\documentclass[11pt]{article}
\usepackage{rotating}

\usepackage{lscape}

\usepackage{graphicx}
\begin{document}
\title{\bf Globular Clusters of the Galaxy: Chemical Composition 
vs Kinematics}

\author{{V.\,A.~Marsakov,  V.\,V.~Koval', M.\,L.~Gozha}\\
{Southern Federal University, Rostov-on-Don, Russia}\\
{e-mail:  marsakov@sfedu.ru, litlevera@rambler.ru, gozha\_marina@mail.ru}}
\date{accepted \ 2019, Astrophysical Bulletin, Vol. 74, No. 4, pp. 403--423}

\maketitle

\begin {abstract}

A comprehensive statistical analysis of the relationship between 
the chemical and spatially kinematic parameters of the globular 
clusters of the Galaxy has been performed. The data of the author’s
compilation catalog contain astrophysical parameters for 157~clusters 
and the relative abundances of $\alpha$-elements for 69~clusters. 
For 121~clusters, the data are supplemented by spatially kinematic 
parameters taken from the literature. The phenomenon of reddening 
of horizontal branches of low-metal accreted globular clusters is 
discussed. We consider the contradiction between the criteria for 
clusters to belong to the subsystems of the thick disk and the 
halo in terms of chemical and kinematic properties. It consists
in the fact that, regardless of belonging to the galactic subsystems 
by kinematics, almost all metallic ($\rm{[Fe/H] > -1.0}$) clusters 
are located close to the center and plane of the Galaxy, while among 
the less metallic of both subsystems there are many distant ones. 
Differences in the abundances of $\alpha$-elements in the stellar objects
of the Galaxy and the surrounding low-mass dwarf satellite galaxies 
confirm the wellknown conclusion that all globular clusters and 
field stars of the accreted halo are remnants of galaxies of
higher mass than the current environment of the Galaxy. A possible 
exception is a distant low-metal cluster with low relative abundance 
of $\alpha$-elements Rup~106.

\end{abstract}


\maketitle

\section{Introduction}

Globular clusters are one of the oldest objects
in the Galaxy, and therefore cause great interest in
connection with the possibility to understand how
the formation and early evolution of the Milky Way
happened. Until recently, all globular clusters were
considered to be typical representatives of their own
galactic halo, that is, formed from a single protogalactic
cloud at the initial stages of the formation
of the Galaxy. However, it was subsequently shown
that some of the clusters most likely landed on our
Galaxy from decaying satellite galaxies. This discovery
roughly coincided in time with the emergence of
the theory that massive galaxies like ours are formed
in the early stages of their evolution as a result of the
continuous accretion of dwarf galaxies. In paper \cite{1}
(hereinafter MKG19) we provide a list containing 22
supposedly accreted clusters, and references to works
in which this is stated. Among them are ten clusters
lost by the dwarf galaxy Sagittarius (Sgr), seven
clusters-by the Canis Major galaxy (CMa) and five
clusters-by other satellite galaxies.

A characteristic feature of such clusters was
the phenomenon of abnormally reddened horizontal
branches, not corresponding to their low metallicity,
and often large distances from the galactic center.
The excess of stars on the red part of the horizontal
branch at low metallicity appears in the case of a
younger age of the cluster, so they were initially
considered ``young''. It is on these grounds that they
were selected when researchers wanted to explore
them in more detail. It is generally accepted that such
clusters form a subsystem, which, according to the
chosen dominant feature, is called the ``young halo'',
``external halo'' or ``accreted halo'' (see, for example,
paper \cite{2} and references therein). But the remaining
genetically related clusters, which are believed to
have formed from a single protogalactic cloud, are
divided into two subsystems: the thick disk and the
halo itself. The separation is due to the metallicity
distribution of the clusters, which reveals a sharp dip
in the vicinity of $\rm{[Fe/H]}\approx -1.0$ (see, for example, the
papers \cite{2},\cite{3}). At the same time, more metal-rich
clusters are considered representatives of the thick
disk subsystem of the Galaxy.

For a long time it was believed that the formation
of all stars in each cluster occurred simultaneously,
and therefore the abundances of all chemical elements
in the stars must correspond to the abundances in the
primary protoclouds of these clusters. But then found
out, that in all clusters the self-enrichment was happening,
which changed abundances of some chemical
elements (see, for example, paper \cite{4} and references
therein). In general, the abundances of only
those chemical elements that are involved in proton
capture processes occurring during hydrostatic helium
combustion in the center or in the layer sources
of asymptotic giant branch (AGB) stars turned out
to be distorted. Mainly, these processes in AGB
stars decrease the relative abundances of primary $\alpha$-
elements (oxygen and to a lesser extent magnesium)
and increased sodium and aluminum. When such a
star loses its envelope at a later stage of evolution,
these elements enter the interstellar medium of the
cluster. As a result, new generations of stars in it are
with a changed chemical composition. The average
abundances of the remaining chemical elements in
cluster stars remain almost primary (see, for example,
paper \cite{5} and references therein). This allows us to
use these abundances to understand the evolution of
the Galaxy in the early stages of its formation.

Since globular clusters belonging to different subsystems
were formed from interstellar matter that has
experienced different scenarios of chemical evolution,
it can be expected that the relative abundances of
chemical elements in clusters of different nature will
differ. This work is devoted to a comparative statistical
analysis of the relationship of the relative abundances
of $\alpha$-elements with the spatially kinematic
characteristics of globular star clusters belonging to
different subsystems of the Galaxy, in order to clarify
their nature and verify the results of our previous
work \cite{1} (MKG19), where the same problem was
investigated for almost half the number of clusters
with known kinematic data, and of lower quality.


\section {INITIAL DATA}

Our catalog was created on the basis of the computer
version of Harris \cite{6} compilation catalog, which
includes all measured values for 157~globular star
clusters of the Galaxy. These data are supplemented
by the relative abundances of 28~chemical elements
in the stars of 69~globular clusters from 101~papers
published from 1986 to 2018. References to these
papers can be found in the on-line catalog, which
is the Appendix in MKG19, which gives a detailed
description of the procedure for averaging the relative
abundances and their errors, and also shows
that the external convergence of the definitions of
the chemical composition of different authors lies in
the range $\langle\sigma\rm{[el/Fe]}\rangle = (0.06-0.16)$. 
Moreover, the
values of the external convergence of the definitions
of the relative abundances of chemical elements in
clusters turned out to be only slightly more than the
variances of the abundances in cluster stars declared
by the authors of the primary papers. This indicates
the absence of significant discrepancies between the
definitions of the abundances by different authors
and the possibility of using our compiled abundances
of chemical elements for statistical analysis of the
chemical composition of clusters belonging to different
subsystems of the Galaxy. Here we consider
the behavior of relative abundances in globular clusters
of only four chemical elements: magnesium,
silicon, calcium and titanium as the most informative
in terms of describing the evolution of the early
Galaxy. Most attention will be paid to the relative
abundances of calcium and titanium. In the visible
spectrum, these two chemical elements have many
lines, and their abundances are quite reliably determined.
The choice of these elements is due to the fact
that the average relative abundances of both primary
$\alpha$-elements --- oxygen and magnesium --- in the course
of cluster evolution decrease compared to their abundances
in primary proto-clouds. And the abundances
of another $\alpha$-element --- silicon —-- are determined for
a smaller number of clusters and are not at all determined
for field stars and dwarf satellite galaxies,
which we use for comparison.

For all 157~globular clusters, we calculated the
rectangular coordinates from positions and distances
from \cite{7}, and for 115 of them-from \cite{8}. We supplemented
the data for the last 115~clusters with
cylindrical velocity components, taking them from
the authors of [8], and for the clusters Ter~4, Pal~3,
Pal~5, Pal~13, NGC6528 and NGC7006 calculated
them from their proper motions, radial velocities and
distances from \cite{7}. As a result, the number of clusters
with known velocities increased by more than two-thirds
compared to our previous work MKG19 and
totals 121~objects. Among them are 63~clusters (45 --
in MKG~19) with the found abundances of chemical
elements. The rectangular velocity components in
a cylindrical coordinate system, the center of which
is placed in the center of the Galaxy, were obtained
on the basis of modern deep surveys (see studies \cite{8},
\cite{9}). Previously, cross-identification was performed
for objects of the catalogs USNO-B1, (U. S. Naval
Observatory A1.0 catalogue), 2MASS (Two Micron
All-Sky Survey), URAT1 (The First U. S. Naval
Observatory (USNO) Astrometric Robotic Telescope
Catalog), ALLWISE (The Wide-field Infrared Survey
Explorer et IPAC), UCAC5 (New Proper Motions
using GaiaDR1) and GaiaDR1 (based on the Tycho-
Gaia Astrometric Solution) with subsequent reduction
to the system Gaia DR1 TGAS (Tycho Gaia
Astrometric Solution). In so doing, the difference
of epochs was up to 65 years, the velocity of the
Sun relative to the galactic center was taken equal
to $(U, V,W)_{\odot} = (-10, 12 + 237, 7)$~km\,s$^{-1}$, and its
galactic distance --- $R_{GC} = 8.3$~kpc. The average internal
error declared by the authors for determining
the components of spatial velocities is approximately
17~km\,s$^{-1}$. Elements of galactic orbits calculated
from these velocities are also taken from \cite{8}. We used
the latter here solely to distinguish clusters that have
orbital points further than 15~kpc from the galactic
center (see below for more details). The ages of
clusters we took from papers \cite{10} and \cite{11}.

To facilitate the understanding of the text and detail
of the figures, we have compiled a Table in which
we have included the parameters of globular clusters
used in this work, as well as indicated the belonging
of clusters to a particular subsystem or group.
The columns of the Table contain the following information:
(1) --- cluster name, (2)--(4) --- heliocentric
coordinates (x, y, z) in the right-hand orthogonal
system in kiloparsecs, (5) --- metallicity $\rm{[Fe/H]}_{H}$ according
to the Harris catalogue. Columns (6)--(8)
contain the iron abundances we found, as well as
the relative abundances of two and four $\alpha$-elements,
further—velocity components in the cylindrical coordinate
system: $V_{R}$, $V_{\Theta}$, $V_{Z}$, where the component
$V_{R}$ is directed to the anti-center of the Galaxy,
$V_{\Theta}$ --- in the direction of galactic rotation, 
$V_{Z}$ --- to the
North pole of the Galaxy (see above for details on
the sources of cluster positions and velocities). This
is followed by Galactocentric distance (12), absolute
magnitude (13), and color index HBR\footnote{Morphological 
index, or horizontal branch color HBR =
(B -- R)/(B + V + R), where B, V, R are respectively the
number of stars at the blue end of the horizontal branch, in
the instability band, and at the red end.}. The last two
columns, (15) and (16), show the clusters’ belonging
to the galactic subsystems: T, TD and H are thin disk,
thick disk and halo, respectively, and the clusters’
belonging to the groups: i -- internal: located at a
distance (or apogalactic radii of their orbits) less than
8 kpc, o -- distant: located at a distance or have radii of
orbits more than 15 kpc, r -- retrograde, with velocity
of azimuthal components less than zero, a -- accreted,
for which the works of other authors prove their extragalactic
origin, t -- genetically related--clusters that
did not fall into any of the last three groups.
For comparison, we used the catalog \cite{12}, which
gives metallicity, relative abundances of $\alpha$-elements
and components of spatial velocities for 785 stars of
the galactic field in the entire range of metallicity of
interest.

\section {STRATIFICATION OF GLOBULAR
CLUSTERS BY GALAXY SUBSYSTEMS} 

In \cite{5} for the first time the stratification of globular
clusters by subsystems of the Galaxy was carried out
using not traditional criteria for metallicity and morphology
of the horizontal branch, but by the components
of their residual velocities, as has been done for
field stars for a long time. In the compilation catalog
for 45 clusters cited in this work, the abundances
of some chemical elements are found, and for 29 of
them there is kinematic information. The authors
found that most of the clusters by kinematics belong
to the galactic halo, however, a significant number of
clusters turned out to have disk kinematics, three of
them--that of thin disk. More than a dozen clusters
were included in the accreted halo, for which, according
to the elements of their galactic orbits, different
authors showed that they were most likely captured
from several satellite galaxies, that is, of extragalactic
origin.

It is clear that a single and sufficient criterion for
stratification of globular clusters by subsystems of the
Galaxy does not exist. To reliably classify a cluster as
a particular subsystem, one should take into account
many parameters characteristic of each subsystem,
in particular, position, kinematics, metallicity, abundances
of different chemical elements, age and morphology
of the horizontal branch. Since we are going
to study the differences in the chemical composition of
clusters of different subsystems, here, as in the previous
work, we used the kinematic criterion in which,
according to the velocity components $V_{R}$, $V_{\Theta}$, $V_{Z}$, the
probabilities of clusters belonging to the subsystems
of a thin disk, a thick disk, and a halo are calculated
by the method described in \cite{13}. This technique is
similar to the technique used in the study \cite{5}, with
only slightly different velocity dispersions in the subsystems.
In both methods, it is assumed that the
components of the spatial velocities of stars in each
subsystem obey normal distributions. As the subsequent
comparison showed, the belonging in the same
clusters turned out to be different only in the case of
differences in the input velocities--here they are more
accurate. As shown by the subsequent comparison,
the belonging of the clusters of the same name turned
out to be different only in the case of differences in the
input velocities--we have them more accurate. Since
belonging to the subsystems are calculated from the
residual velocities, we have reduced the azimuthal
components of the cluster velocities to the rotation
velocity of the centroid at the galactocentric distance
at which the cluster is located. We took the rotation
curve from the Galaxy model \cite{8}. Taking into account
the large distance of clusters, which leads to large
errors in determining tangential velocities, and also
taking into account that clusters do not concentrate
to the galactic plane and do not participate in the
general rotation of the galactic disk, as close stars
of the field, for which the \cite{13} method was developed,
we performed a recurrent procedure when calculating
the probabilities of falling clusters in a particular subsystem.
In the second step, we assigned the velocity
dispersions and the numbers of clusters in subsystems
in the probability formulas such values as we
obtained after the first step. This reduced the specified
portion of objects in the subsystems of the thin and
thick disks. Although the recalculation somewhat
redistributed the belonging of a number of clusters
that are kinematically in the transition zones between
the thin and thick disks, as well as between the thick
disk and the halo, but in general the composition of
subsystems has changed insignificantly. To eliminate
the ambiguity of individual stratification in paper \cite{13}
it is recommended to consider a star as belonging to
a subsystem if the probability of its belonging to an
alternative subsystem is at least two times less. As a
result of this assumption, there is no stratification for
a number of sample stars even with kinematic data.
Such stars are usually referred to as intermediate. In
view of the fact that we are interested in statistical
regularities in subsystems, which can be identified
only by a significant number of objects, we stratified
all our globular clusters by subsystems, believing the
sufficient criterion is simply a higher probability of
belonging to any subsystem. The Table indicates the
belonging to the galactic subsystems for those clusters,
for which we have estimates of spatial velocities.

Fig. 1a presents the Tumre diagram 
``$V^{2}_{\Theta}$ -- $(U^{2}_{R} + W^{2}_{Z})^{0.5}$'' 
for our globular clusters and field stars from
the study \cite{12}. The figure shows that objects demonstrating
the kinematics of the subsystems of the same
name occupy approximately the same areas on the
diagram, although in \cite{12} the technique slightly different
from our one was used for field stars. The
application of our technique showed that, from the
kinematic parameters, for 84 clusters (40 in MKG~19)
the probability of belonging to the halo is greater than
to other subsystems. 34 clusters (28 in MKG~19)
more likely belong to the thick disk, and three clusters
(four in MKG19) turned out to be with the kinematics
of the thin disk. We see that another model of the
rotation curve of the Galaxy, more accurate estimates
of the velocity, as well as a significantly increased
number of velocity components, slightly reduced the
number of clusters in the thin disk, but the population
of the thick disk increased by almost a quarter, and
the number of halo clusters more than doubled. Note
that a comparison of the velocity components showed
a noticeable difference for the correlation coefficients
of the same-named components $r = (0.6-0.7)$. And
the differences in velocities, as shown by the test,
are caused mainly by the refinement of their proper
motions, and not the distances to the clusters. As a
result, the subsystems did not have the same clusters
in all cases. In particular, none of the three clusters
with the thin disk kinematics (according to the
results of this work) belonged to this subsystem in
MKG19. In the thick disk, 17 clusters were common.
Estimates of the velocities of about a dozen
clusters have changed so that instead of a thick disk,
as in the previous work, in the present work they
fell into the halo. Significant differences in velocities
should be noted. So, according to the data on
velocities from paper \cite{7} used in our previous work,
the greatest velocity was for the cluster NGC~6553—
$V_{\Theta} = 383$~km\,s$^{-1}$ and now this value is 176~km\,s$^{-1}$.
The azimuthal components of the velocities of two
other clusters --- Pal 6 and NGC~6284 --- decreased as
much, and for three clusters --- NGC~4147, NGC~6144
and NGC~6723 --- they increased. In Fig.~2a, it can
be seen that among the clusters with the kinematics
of the thick disk there was a significant number
of such objects with rotational velocities around the
galactic center even larger than those of the Sun. Two
thick disk clusters (NGC~6496 and NGC~6528) have
azimuthal velocities above 300~km\,s$^{-1}$ at all. Four
clusters with such large direct azimuthal velocities
are also present in the halo. It is also seen that
more than half of the halo clusters exhibit retrograde
rotation around the galactic center. We believe that
such clusters are highly likely to be of extragalactic
origin. Indeed, according to the hypothesis of the
monolithic collapse of the protogalaxy from the halo
to the disk \cite{14}, the field stars and globular clusters
genetically associated with the Galaxy cannot be in
retrograde orbits.

Fig.~1b shows the distribution of clusters in the
coordinates ``the distance from the galactic plane
(z) -- the metallicity [Fe/H]''. In diagrams where spectroscopic
determinations of other chemical elements
are not used, the [Fe/H] values were taken from the
catalog \cite{6}, since it contains the metallicities for all
clusters. Large circles in the figure indicate clusters
belonging, according to kinematic characteristics, to
a thin disk (empty), a thick disk (red) and a halo
(dark gray), and asterisk are unstratified clusters,
that is, clusters with unknown velocities. The most
noticeable detail in the figure is the high concentration
of metallic ($\rm{[Fe/H]} > -1.0$) clusters near the
galactic plane, regardless of whether it belongs to
the subsystem of the Galaxy determined by kinematic
criteria. As the test showed, the maximum distance
of the orbit points of all\footnote{The exception is three
 distant metal clusters Pal 12, Whiting
1 and Terzan 7, which most likely belonged to the last
decayed dwarf galaxy Sagittarius.} metal clusters were less
5~kpc, while for a significant part of small metal
clusters $Z_{max} > 10$~kpc. This fact, along with a
pronounced dip on the metallicity function in the
region $\rm{[Fe/H]} \approx -1.0$ also encourages the allocation
of metal-rich clusters in the disk subsystem. But, on
the other hand, the figure shows that the vast majority
of clusters with thick disk kinematics demonstrates
$\rm{[Fe/H]} < -1.0$, which is in contradiction with the
above-described practice of allocating disk clusters
by metallicity. The concentration of metallic clusters
to the galactic plane forms a long-known vertical
metallicity gradient. The situation is similar with
the radial metallicity gradient. In Fig.~1b, it can be
seen that at a distance of $z \geq 5$~kpc, the main part of
the clusters lost by satellite galaxies is located. And
in such clusters, the spatial velocities reflect not the
dynamic conditions of star formation in a shrinking
protogalactic cloud, but only the finite orbits of clusters
captured from dwarf satellite galaxies that have
disintegrated under the tidal forces of the Galaxy. At
the same time, the more massive the parent satellite
galaxy, the more flat and elongated its orbit, at which
it loses its clusters and stars \cite{15}.

\section {PROPERTIES OF GLOBULAR CLUSTERS
OF DIFFERENT SUBSYSTEMS
AND WITH DIFFERENT NATURES}

Fig. 2a shows the diagram ``azimuthal velocity
$V_{\Theta}$ -- metallicity [Fe/H]'' for globular clusters and field
stars. Different icons indicate objects of different
subsystems of the Galaxy. In contrast to a similar
diagram in \cite{5}, this one shows clusters with azimuthal
velocities significantly different from the solar one in
the range $\rm{[Fe/H]} > -1$. Moreover, four metal-rich
clusters with retrograde orbits are within 2~kpc from
the galactic center, and one--at 4~kpc, whereas 25
metal-poor clusters with $V_{\Theta} < 0$ have an average
galactocentric distance of about 8~kpc (15 clusters
lie further than 4~kpc). In the diagram ``$V_{\Theta}$ -- [Fe/H]'',
we have also identified clusters that at different times
various authors have considered to belong to formerly
decayed dwarf satellite galaxies. Additionally, clusters
with galactocentric distances $R_{G}$ or maximum radii
of orbits $R_{max}$ above 15~kpc are noted. As you can
see, only one accreted cluster ($\omega$~Cen) is not marked,
that is, it is closer than this radius. While fifteen other
clusters lie far away, their extragalactic origin is not
proven. The behavior diagram of ``distant'' clusters
does not agree with the analogous one in MGK19,
according to which there is a significant correlation
between metallicity and the azimuthal component of
velocity. In this work, there are more distant clusters
due to the increase in the spatial velocity components
and elements of galactic orbits, and there are many
distant metal-poor clusters with high azimuthal velocity
components. As a result, the correlation noted
earlier is disavowed here.

The diagram ``morphology of the horizontal branch
HBR -- metallicity'' for our clusters is shown in Fig.~2b.
It follows that most of the clusters (but not all) that
are currently inside the solar circle ($R_{GC} < 8$~kpc)
actually have mostly extremely red or extremely blue
horizontal branches. But between these extreme
positions, part of the inner clusters lies in a thin layer
along the upper envelope in the diagram, while most
of the known accreted clusters lie mostly below it (see
the inclined line in Fig.~2b drawn ``by eye''). However,
as can be seen in the diagram, this arrangement is
not absolute, and there are exceptions. The probable
accreted clusters are most likely also distant clusters
($R_{GC}$ or $R_{max} > 15$~kpc) and clusters with retrograde
rotation ($V_{\Theta} < 0.0$). Thirteen of the 26 retrograde
clusters lie inside the solar circle. Nine retrograde
clusters have extremely blue horizontal branches, five
are extremely red, and for ten, branches turned out to
be abnormally reddened for their low metallicity. As
can be seen in Fig.~2b, all reddened branches really
lie in the range between the extreme HBR values in
the diagram below the upper envelope. And among
the distant clusters all (except three) were metal-poor,
whereas the morphology of their horizontal
branch can be any. It is generally believed that all
metal-poor clusters lying below the narrow upper
band can with a high probability be considered as
candidates for accreted (see \cite{16}). It seems that
this assumption is very plausible, and the redness of
the horizontal branches of accreted clusters can be
explained. Information on the belonging of clusters to
the groups mentioned above is given in the Table.

As we have already noted, recent studies show
that within the most massive globular clusters there
are several episodes of star formation with supernova
explosions that enrich the interstellar medium
of the cluster with elements of the iron group. For
example, several populations differing in metallicity
are found in the largest cluster of Omega Centauri
($\omega$~Cen). However, star populations differing
in the abundance of helium and CNO elements are
also found in less massive clusters (see, for example,
paper \cite{17}). It is believed that new, younger star
populations in such clusters are formed from chemically
contaminated matter ejected by giants of the
asymptotic branch of intermediate masses, rapidly
rotating massive stars, and also rotating AGB stars
of the first generation \cite{18},\cite{19}. Over time, extended
horizontal branches form in such clusters, as a result
of which their color ceases to correspond to the primary 
metal-poor chemical composition of the stellar
population dominant in number. The authors of \cite{20}
showed by numerical modeling that in clusters with a
secondary younger population enriched mainly with
CNO-elements, the color of the horizontal branch
actually becomes more red. At the same time, the
Oosterhoff type of cluster changes. As the simulation
shows, all this occurs in the initial stages of the
cluster's evolution within one billion years after the
last burst of star formation.

In our diagram in Fig.~2b, we can see that of
42 clusters with extremely blue horizontal branches
($\rm{HBR} > 0.85$), only eleven are located or have orbit
points further than 15~kpc from the galactic center,
while 29 clusters are currently inside the solar circle
($R_{GC} < 8$~kpc), and four clusters are between these
boundaries. Moreover, among distant clusters, almost
all are rather weak in absolute magnitude, that
is, their masses are small. This is well illustrated in
Fig.~2c, where the diagram ``distance from the center
of the Galaxy $R_{G}$ -- the absolute magnitude $M_{V}$'' for
clusters with extremely blue horizontal branches is
shown. In addition, the figure highlights distant clusters
($R_{GC}$ or $R_{max} > 15$~kpc). We see that all such
clusters have $M_{V} \geq 8^{m}.0$ (the exceptions are the
distant bright cluster NGC~2419 from Sgr, which has
a boundary value HBR = 0.86, and NGC~4833 with
a boundary luminosity $-8^{m}.16$), while all brighter
clusters turned out to be close. However, the predominance
of low-mass clusters among distant clusters
has long been known (see article \cite{21} and references
therein), but for extremely blue clusters this regularity
is most pronounced. As a result, it turns out
that in low-metal clusters, extremely blue horizontal
branches are mainly observed near the galactic center
and in a small number of distant, relatively low-mass
clusters. The reason may be that in both types of
clusters, the matter ejected by evolved stars does not
remain in the clusters, but is swept away by disturbances
of the gravitational potential of the Galaxy.
Moreover, in the former, this is due to frequent approachings
to the galactic bulge and disk, while in
the latter--because of their small mass, unable to
hold this matter even at a considerable distance from
the galactic center. As a result, the secondary population
of them either does not form, or is formed in
a small amount. In metal-poor clusters with reddened
horizontal branches, all points of the orbits are
often outside the solar circle, where disturbances of
the gravitational potential of the Galaxy affect less.
Perhaps that's why they have time to form a population
of younger stars, distorting the color of their
horizontal branches. The described picture is not
quite unambiguous, since the third and subsequent
populations in some clusters are over-enriched with
helium, which leads to the appearance of stars on the
horizontal branch on the high-temperature side of the
instability band. As a result, the color of the branch
shifts to the blue side. A typical example is the cluster
M15, which in addition to the normal blue part of
the horizontal branch also has the so-called ``blue
tail'' \cite{22}. Verification of the proposed explanation for
the existence of a correlation between the color of the
horizontal branch and the loss of gas by the cluster
requires a detailed analysis of the orbital tracks of the
clusters, as well as the use of numerous published
data on the individual chemical composition of stars
in clusters.

Fig.~2d shows the diagrams of the ``azimuthal
velocity $V_{\Theta}$ -- ratios [Ca,Ti/Fe]'' of globular clusters
of our sample, for which these parameters are determined,
and field stars. Additionally, clusters for
which extragalactic origin, that is, considered to be
accreted, and distant clusters are indicated on the
diagram. The vertical line $V_{\Theta} = 0$ separates field stars
and clusters with retrograde rotation. Field stars are
characterized by high (on average) ratios [$\alpha$/Fe], but
with a large spread at small and negative values of
the azimuthal velocity and their rapid decrease near
the velocity of rotation of the galactic disk at a solar
galactocentric distance. In globular clusters with any
kinematics, the relations [$\alpha$/Fe] differ little from each
other and do not correlate at all with the azimuthal
component of the velocity. And at all values 
$V_{\Theta}<V_{\odot}$ their dispersion 
is small ($\sigma\rm{[\alpha/Fe]} \approx 0.1$), however at
the velocity of revolution around the galactic center
more than solar the dispersion sharply increases (the
truth, such clusters are only six, but among them
there are clusters both with kinematics of all three
subsystems, and clusters of various origin). The
high relative abundances of $\alpha$-elements suggest that
almost all clusters were formed, most likely, from
interstellar matter not yet enriched with elements of
the iron group from type Ia supernova outbursts.

\section {ABUNDANCES OF $\alpha$-ELEMENTS IN
GLOBULAR CLUSTERS OF DIFFERENT
SUBSYSTEMS AND ORIGIN}

Fig.~3a shows the diagram ``[Fe/H] -- [Ca,Ti/Fe]''
for globular clusters of different galactic subsystems
and field stars of different nature (details below). The
figure shows that clusters belonging to any subsystem
by the kinematic criterion, unlike field stars, can
have not only different metallicities, but also different
relative abundances of $\alpha$-elements. In Fig.~3b,
where the same diagram is plotted for averaged over
four $\alpha$-elements: magnesium, silicon, calcium and
titanium, we see that in general the position of the
region occupied by globular clusters relative to field
stars has not changed, but the number of clusters
decreased. Unlike other diagrams of the figure, on
this panel, for comparison, different icons indicate
field stars of different galactic subsystems, selected by
the kinematic criterion from \cite{13}. It can be seen that
the clusters and field stars of the same-name subsystems
have a significantly different chemical composition.
For genetically connected field stars, that is,
formed from a single protogalactic cloud, metallicity
can serve as a statistical indicator of their age, since
in a closed star-gas system (which in the first approximation
can be considered our Galaxy), the total
abundance of heavy elements steadily increases over
time. As such, we consider field stars with residual
velocities $V_{res} > 240$~km\,s$^{-1}$ (see study \cite{23}), and they
are designated by small dark green snowflakes on
the diagram. The vast majority of field stars with
higher residual velocities (indicated by crosses) have
retrograde rotation (see Fig.~2a). All such high-velocity
stars can be considered candidates for accreted.
Note that metal-poor ($\rm{[Fe/H]} < -1.0$) genetically
related field stars are located along the upper
half of the strip in Fig.~3a, 3c, 3d. To orient in the
figures the broken curves, drawn ``by eye'', indicate
the lower envelopes for genetically related field stars.
The position of our line is in good agreement with
paper \cite{24}, the authors of which distinguished two
populations among metal-poor field stars not by kinematics,
but by the relative abundances of $\alpha$-elements
approximately along the boundary $\rm{[\alpha/Fe]} =\sim0.3$ and
found that the stars of these populations differ not only
in chemical composition, but also in kinematics and
age. Moreover, they initially looked for evidence that
the population with a lower relative abundance of 
$\alpha$-elements is of extragalactic origin. Figs.~3a, 3c, 3d
show that in genetically related stars, the [$\alpha$/Fe]
ratios sharply decrease with increasing metallicity,
starting from $\rm{[Fe/H]} \approx -1.0$, due to the onset of Ia
SNe outbursts in the Galaxy. In globular clusters,
this is not observed, and the vast majority of metal-rich
clusters lie above the band occupied by field stars.
Although a slight decrease in the [$\alpha$/Fe] ratios with
an increase in metallicity in the range $\rm{[Fe/H]} > -1.0$,
is noticeable for them, their positions in the diagram
mainly remain in the ratio range $\rm{[\alpha/Fe]} > 0.15$, as in
less metallic clusters. In this case, clusters belonging
by the kinematics to two most numerous galactic
subsystems--thick disk and halo--do not show statistically
significant differences in positions in Figs.~3a
and 3b.

Fig.~3c shows the same ``[Fe/H] -- [Ca,Ti/Fe]'' diagram,
but the clusters are distinguished by other
signs: accreted clusters, whose membership in the
past to decayed satellite galaxies was determined by
different authors based on estimates of their positions
and spatial movements, distant clusters ($R_{GC}$
or $R_{max} > 15$~kpc) and retrograde ones. It can be
seen that in the range ($\rm{[Fe/H]} < -1.0$), clusters are
located on the diagram so that the lower envelope for
genetically related stars is close to the median. It is
also seen that, in general, the entire population of accreted
clusters, together with candidates for accreted
clusters (distant clusters and clusters in retrograde
orbits), shows in Fig.~3c a large scatter of the [$\alpha$/Fe]
ratios. (Moreover, five of the twelve clusters with
retrograde orbits were inside the solar circle.) Their
spread is much larger than that of genetically related
field stars. However, approximately the same wide
spread is shown in Fig.~3 by high-velocity ($V_{res} >
240$~km\,s$^{-1}$) metal-poor field stars that are not genetically
related to a single protogalactic cloud and are
likely extragalactic in origin. It is possible that a large
spread of the [$\alpha$/Fe] ratios in such clusters and field
stars could be due to the difference in the maximum
masses of type II supernovae that enriched the matter
of their many parent dwarf galaxies.

In Fig.~3d, solid black circles show clusters that,
by any signs, cannot be classified as candidates for
accreted ones. We consider such clusters to be genetically
related, that is, formed from a single protogalactic
cloud. By definition, all 32 such clusters of our
sample are located closer than 15 kpc from the galactic
center. Moreover, 27 of them, marked in the figure
by white triangles inside the circle, generally lie inside
the solar circle ($R_{GC} < 8$~kpc). And among them
are all metal-rich clusters with high ratios [Ca,Ti/Fe],
some of which most likely belong to the galactic bulge
(see \cite{25}). In this diagram, in addition to genetically
related clusters, those are also plotted for which
belonging to two very massive dwarf galaxies--Sgr
and CMa--is considered to be reliably established
(see the list of accreted clusters in MKG19). The
figure shows that 21 out of 27 accreted and genetically
related clusters in the range $\rm{[Fe/H]} < -1.0$ form a
rather narrow band on the diagram, and the lower
envelope for genetically related field stars can serve
as such for them also. But at the same time, all
of these clusters are more closely concentrated to
this line than genetically related field stars. (The
very low [Ca,Ti/Fe] ratios are shown by two metal-poor
clusters from Sgr: a very distant cluster with
the only star studied in one work NGC~2419 and
Ter~8, however, the relative magnesium abundances
in both are high --- 0.30 and 0.52, respectively, and
the silicon abundance in Ter~8 is 0.38, that is, when
all $\alpha$-elements are taken into account, these clusters
also appear near the lower envelope. The situation
is similar with the halo cluster NGC~6287, for which
[Si/Fe] = 0.55, a [Mg/Fe] = 0.35.) But in a more
metallic range, both metal-rich clusters (Pal~12 and
Ter~7) captured from a dwarf satellite galaxy of Sgr
lie below field stars. The metal-poor cluster Rup~106,
which is believed to be lost by the rather massive
dwarf galaxy CMa \cite{26}, is also anomalously low in
the diagram. However, the very low relative abundance
of $\alpha$-elements in it along with low metallicity
contradicts this assumption. It can be assumed that
it was lost by one of the low-mass dwarf satellite
galaxies. Provided, of course, that the abundances of
$\alpha$-elements in just two stars of Rup~106 are defined
correctly in one article. Note that this cluster is
one of the least massive ($M_{V} = -6^{m}.35$) metal-poor
clusters in the Galaxy and could well be formed in
such a dwarf galaxy.

\section {ACCRETED GLOBULAR CLUSTERS
AND MASSES OF THEIR PARENT
GALAXIES} 

In \cite{27}, for 235 stars of the core of the currently
destroyed dwarf galaxy Sagittarius (Sgr), the dependence
of [Mg, Ca/Fe] on [Fe/H] is constructed
and it is emphasized that in the low-metal range
($\rm{[Fe/H]} < -1.0$) the sequence of stars from this
galaxy coincides with the sequence of field stars of
the Galaxy, and with greater metallicity it lies slightly
lower than that of field stars. Moreover, the authors
of this work note that in the region $\rm{[Fe/H]} > -1.0$ the
dependence of the relative abundances of $\alpha$-elements
on metallicity in the Sagittarius galaxy is very similar
to that observed in stars of the most massive satellite
of the Galaxy--Big Magellanic Cloud. This, in their
opinion, also implies a large mass of the Sagittarius
galaxy. Indeed, the simulation in \cite{28} of the kinematics
of the tidal tail of the stars of the Sagittarius
galaxy showed that in order to reproduce the velocity
dispersion in the stream from this galaxy, the mass
of its dark halo should be $M = 6 \times 10^{10} M_{\odot}$. The
authors of \cite{27} managed to reproduce the observed
chemical laws in the parent dwarf galaxy Sagittarius
in a model that implies just such a large initial mass
and a significant loss of it several billion years ago,
starting from its first intersection of the perigalactic
of our Galaxy.

In Fig.~3d, among the field stars, the field stars
of the so-called Centaurus stream are marked with
large gray oblique crosses. It is assumed that all
these stars were lost by a dwarf satellite galaxy, the
central core of which was the most massive globular
cluster Omega Centauri ($\omega$~Cen) currently owned by
our Galaxy (see article \cite{29} and references therein).
Numerical simulation of the dynamic processes occurring
during the interaction of the satellite galaxy
with the disk and bulge of our Galaxy showed that
the capture of the dwarf galaxy core into an elongated
retrograde orbit with a small apogalactic radius
is quite possible, while the galaxy should be
quite massive: of order of 10$^{9}M_{\odot}$ \cite{30}. In particular,
the results of numerical simulation \cite{31} demonstrated
that the orbits of sufficiently massive satellite galaxies
are constantly decreasing in size and moving into
the galactic plane by dynamic friction. Over time,
such galaxies, having acquired very eccentric orbits,
almost parallel to the galactic disk, begin to be intensively
destroyed by the tidal forces of the Galaxy
with each passing perigalactic distance, losing stars
with clearly determined orbital energies and angular
moments. Therefore, if the observer is between
the apogalactic and perigalactic radii of such an orbit,
the tidal ``tail'' from the disrupted galaxy will be
observed as a ``moving group'' of stars with small
vertical velocity components and a wide, symmetrical
and often bimodal distribution of radial spatial
velocity components. Based on the recommendations
of the authors of paper \cite{32}, in paper \cite{23} from
the author's cumulative catalog of spectroscopic determinations
of magnesium abundances (representative
of $\alpha$-elements) in 800 close F\,--K\,--dwarfs of
the field \cite{33} by the azimuthal and vertical velocity
components in the ranges $-50 \leq V_{\Theta} \leq 0$~km\,s$^{-1}$ and
$|V_{Z}| < 65$~km\,s$^{-1}$, respectively, stars that were lost by
a dwarf galaxy centered on the cluster $\omega$~Cen were
allocated. It turned out that the isolated 18 stars of
the stream did show a rather narrow dependence of
[Mg/Fe] on [Fe/H], characteristic of genetically related
stars. Moreover, the position of the ``knee point''
of the relative magnesium abundance at $\rm{[Fe/H]}\approx -1.3$~dex 
indicates that the rate of star formation in
their parent galaxy was lower than in our Galaxy.
Star formation in this galaxy seems to have lasted so
long that its most metal-rich stars reached a ratio of
$\rm{[Mg/Fe]} < 0.0$~dex, that is, even less than that of the
Sun. However, the low value of the maximum metallicity
of the stars of this group (only $\rm{[Fe/H]} \approx -0.7$)
indicates the cessation of subsequent star formation
in their parent galaxy. Most likely, this was due to
the beginning of the collapse of the dwarf galaxy. In
other words, the chemical composition of the stars
of this former galaxy suggests that it did evolve for
quite some time (but less than our Galaxy) before
collapsing. In this paper, by the same criteria, we
isolated the Centauri stream stars from the field star
catalog \cite{12} used here. There were also 18 such stars
(see Fig. 3d). We see that the behavior of two other
$\alpha$-elements --- calcium and titanium --- corresponds to
the description of the behavior of magnesium according
to another catalog. As a result, it turns
out that the dependence of [$\alpha$/Fe] on [Fe/H] in the
stars of the Centauri stream is in good agreement
with the dependence of accreted clusters in the range
$\rm{[Fe/H]} > -1.5$. That is, the assumption is confirmed
about the extragalactic origin of at least some high-velocity
field stars, that came to us from satellite
galaxies of rather large masses .

\section {DISCUSSION}

OurGalaxy has a complex multicomponent structure,
consisting of several subsystems, which are laid
out in each other. There are no clear boundaries
for subsystems, so their size can be estimated only
approximately. Geometric boundaries imply certain
dispersions of the velocities of objects belonging to
this subsystem. The use of kinematic parameters is
considered the most reliable method of stratification
of objects by subsystems. It is in this way that the
field stars are divided into subsystems of the Galaxy.
As the results of this work show, this method is not
suitable for globular clusters, since the clusters of
different subsystems isolated by kinematics demonstrate
chemical properties that are radically different
from the properties of the field stars of the same
galactic subsystems. In particular, all metal-rich
($\rm{[Fe/H]} > -1.0$) clusters belonging to any subsystem
by kinematics are enclosed within rather narrow
limits relative to the center and plane of the Galaxy.
But in the less metallic range, among the clusters,
isolated by kinematics, of both the thick disk and the
halo, there are also very distant ones. This is evident
in the well-known radial and vertical metallicity gradients
in the general population of globular clusters
of the Galaxy. It turns out that the traditionally used
procedure for isolating the clusters of the thick disk
and halo by metallicity is more acceptable. We emphasize
that the same contradiction was revealed in
our previous work, where the clusters were stratified
by velocities from \cite{7}. It turns out that increasing the
number and reliability of determining the components
of cluster velocities did not lead to its elimination.
The difference in the chemical composition of globular
clusters and field stars, allocated by the kinematic
criterion in the subsystem of the thick disk, indicates,
most likely, the lack of correspondence between the
subsystems of the same name for these objects. It
turns out that the reasons for the formation of these
subsystems are different for field stars and globular
clusters. Indeed, the high relative abundances of
$\alpha$-elements in metal-rich clusters suggest that they
formed within about a billion years after the beginning
of star formation. While in the field stars, starting
from [Fe/H] = 1.0, these ratios begin to decrease due
to the beginning of the era of mass type Ia supernova
outbursts. This suggests that metal-rich field stars
are younger than globular clusters of the same metallicity.

However, we note that a similar contradiction,
although not so pronounced, between the criteria for
belonging to the disk subsystems and the halo by
chemical and kinematic properties is also observed in
field stars of the RR Lyrae type (see paper \cite{34}). Perhaps
not all field lyrids are genetically related to our
Galaxy. In \cite{35} it is proved that some of the field stars
got into it from a rather massive captured satellite
galaxy shortly after the formation of its own halo (see
below for details). By now, some of them may well
have become variables of type RR Lyrae. This version
is in favor of the fact that the relative abundances of
$\alpha$-elements in the field lyrids undergo a kink in the
diagram ``[Fe/H] -- [$\alpha$/Fe]'' at a lower metallicity than
field stars. This version is also confirmed by the somewhat
lower relative abundances of all $\alpha$-elements (the
above is especially true for titanium) in most metal-rich
lyrids with thin disk kinematics (see Figs.~2a -- 2d 
and 3a in work \cite{34}). But then there is a need to
explain how such metal-rich stars could have formed
at the early stages of evolution in a dwarf galaxy now
already, and could have acquired the kinematics of
field stars of the thick and thin disk in our Galaxy.
Of course, this assumption is very superficial and
requires comprehensive justification. 
Note that in \cite{36} we assumed that
relatively young metal-rich field lyrids have increased
helium content, leading to faster evolution of stars,
and in the vicinity of the Sun they are carried out
by radial migration from the central regions of the
Galaxy, where such stars have already been found.

Recall that the probabilities of clusters belonging
to galactic subsystems are calculated on the basis of
residual velocities at galactocentric distances corresponding
to current positions. In so doing, it is not
taken into account at all how far from the galactic
plane the clusters are now. As a result, the vertical
components of the residual velocities of clusters located
near their apogalactic radii of orbits are underestimated.
And this, in turn, can lead to erroneous
attribution of such clusters to the disk subsystem.
However, our verification showed that, even if clusters
located far from the galactic plane were removed from
the clusters of the the thick disk selected by kinematics,
there would still be many metal-poor ones in the
subsystem. Moreover, even the maximum distances
from the galactic disk in the remaining clusters turned
out to be less than 3~kpc. And on the other hand,
among metal-rich clusters, there are still many such
objects with the kinematics of the halo. That is,
the discrepancy between the kinematic and chemical
stratification criteria remains.

If we assume that all genetically related globular
clusters were formed from the matter of a single protogalactic
cloud, then we can assume that the existence
of active phases in the evolution of the Galaxy is
responsible for such a distinguished position of metal-rich
clusters (see the study \cite{3}). The active phase
period occurs after massive supernova bursts in the
halo, warming up interstellar matter, resulting in a
delay in star formation. During this delay, the interstellar
matter of the proto-galaxy, already contaminated
with heavy elements, mixes, cools and collapses
to a smaller size, after which star formation begins
again in the Galaxy and disk subsystems are formed.
However, in such a scenario, the formation of subsystems
in globular clusters does not fit: as can be seen
in Fig.~3a--3d, relative abundances of $\alpha$-elements in
almost all studied metal-rich clusters (except three
accreted clusters Ter 7, Pal 12 and Rup 106, and two
bulge clusters NGC~6528 and NGC~6553) turned out
to be high: $\rm{[\alpha/Fe]} > 0.15$. The absence of a reliably
traceable ``knee'' on the dependence of [$\alpha$/Fe] on
[Fe/H], as in the field stars, indicates that all the studied
clusters were formed before the onset of SNe Ia
outbursts, that is, during the first billion years after
the start of star formation in the protogalactic cloud.
These supernovae enrich the interstellar medium exclusively
with atoms of elements of the iron group, as
a result of which the [$\alpha$/Fe] ratios in the closed star-gas
system begin to decrease. As can be determined
by the field stars in Fig. 3a -- 3d, in the Galaxy this
happens at $\rm{[Fe/H]}\approx -1.0$. The same figure shows
that within the metallic range, clusters also have a
decrease in the relative abundances of $\alpha$-elements
with an increase in metallicity, but the ratios [$\alpha$/Fe]
for any metallicity remains higher than that of field
stars of the thick disk . As a result, the dependence of
[$\alpha$/Fe] on [Fe/H] lies above and parallel to the similar
dependence of the field stars. And among them there
are clusters of all subsystems allocated by kinematics.
As seen in Fig.~3d, what unites the metal-rich clusters
is that they all lie inside the solar circle. And even
the most remote points of their orbits practically do
not go beyond this radius. The estimation of the age of
clusters due to its uncertainty does not allow to draw
a final conclusion about their nature. In particular,
according to estimates of ages from paper \cite{10}, they
are all younger than 11.5~Gyr. But according to the
definitions from paper \cite{11} they are older and appeared
simultaneously with the oldest, least metallic clusters.
On the other hand, both data show a steady
monotonous decrease in the ages of metal-rich clusters
with increasing metallicity. It turns out that more
metal-rich clusters with lower relative abundances of
$\alpha$-elements are born mainly later. In this case, we
have to agree with the statement that this is due to the
beginning of SNe Ia supernovae outbreaks. But then
the small ages of metallic clusters from the study \cite{10}
seem to be more correct. This behavior of clusters
can be put within the framework of the hypothesis
about the active phases of Galaxy evolution. Moreover,
the higher relative abundances of $\alpha$-elements
for all metal-rich clusters than those of the stars
fields, can be explained by the fact that they formed
from the interstellar matter, which already collapsed
and was enriched with heavy elements after delayed
star formation. This interstellar matter of increased
density led to an increase in the upper the mass limit
of the forming stars, and therefore of the supernovae
of the second type, throwing out a greater number
of $\alpha$-elements. In this case, metal-poor, genetically
related clusters, the orbits of which also lie almost
completely inside the solar circle, should have been
born before the onset of the active phase. Indeed,
according to \cite{10}, the ages of less metallic genetically
related clusters are systematically larger. It is possible
that already at the time of their birth, the collapse rate
of the protogalactic cloud had slowed significantly,
which led to the presence of clusters with a ``younger''
kinematics of the thick disk among them. However,
large differences in cluster velocities, according
to estimates of various authors, do not exclude the
possibility that their errors are significantly underestimated,
which could lead to incorrect stratification
of some clusters. Thus, based on the hypothesis
about the active phases of the evolution of the Galaxy,
we can try to give a consistent explanation of the
cause of the jump-like change in the volume in the
Galaxy occupied by clusters when passing through
$\rm{[Fe/H]} \approx -1.0$. Although in this case it is still unclear
the existence of clusters with halo kinematics in the
range $\rm{[Fe/H]} > -1.0$, the appearance of which can
not be attributed to small modern values of errors in
the measurement of proper motions and distances to
globular clusters.

In our Fig.~3 it can be seen that the entire
set of metal-poor ($\rm{[Fe/H]} < -1.0$) globular clusters
of the Galaxy occupies on the ``[Fe/H] -- [$\alpha$/Fe]''
diagram almost the same band along with fast
($V_{\Theta} > 240$~km\,s$^{-1}$), that is, accreted field stars.
Moreover, as follows from the same diagram, the
stars of dwarf satellite galaxies [37.39] of our Galaxy,
with the same low metallicity, have significantly lower
values [$\alpha$/Fe]. The above indicates that all stellar
objects of the accreted halo are remnants of galaxies
of higher mass than the current environment of the
Galaxy. Differences in the abundances of $\alpha$-elements
between the stellar objects of the Galaxy and the
less massive dwarf satellite galaxies surrounding it
indicate that the latter did not leave a noticeable
stellar trace in it. This conclusion is consistent with
the result obtained from a smaller number of globular
clusters in \cite{5}. In a recent article \cite{40}, based on the
discovery of high radial anisotropy of the velocity field
in a large sample of halo dwarfs from a vicinity of
about 10~kpc from the Sun, it was also concluded
that a large-mass satellite accretes to the Galaxy
about 8 -- 11~Gyr ago. Even more definite conclusions
regarding the capture of a massive ($10^{9}M_{\odot}$) satellite
galaxy by our Galaxy at the early stages of evolution
and the formation of a thick disk as a result of its fall
due to heating of an already formed thin stellar disk
were made in \cite{35}. These conclusions were obtained
by analyzing the relative abundances of $\alpha$-elements
and the velocities of several tens of thousands of stars
within 15~kpc from the Sun in a sample compiled
by cross-identification between the catalogs SDSS--APOGEEDR14
and GaiaDR2.

\section*{ACKNOWLEDGMENTS}

The authors thank Alexander Chemel for providing
spatial velocity components for 115 globular clusters
and the rotation curve of his model of the Galaxy.

\section*{FUNDING}

M. V. A. and G. M. L. thank for the support of
the Ministry of Education and Science of the Russian
Federation (state assignment No. 3.5602.2017/BCh),
and K.V.V. thanks for the support of the Ministry of
Education and Science of the Russian Federation
(state assignment No. 3.858.2017/4.6).

\renewcommand{\refname}{REFERENCES}

\newpage

\begin{figure*}
\centering
\includegraphics[angle=0,width=0.99\textwidth,clip]{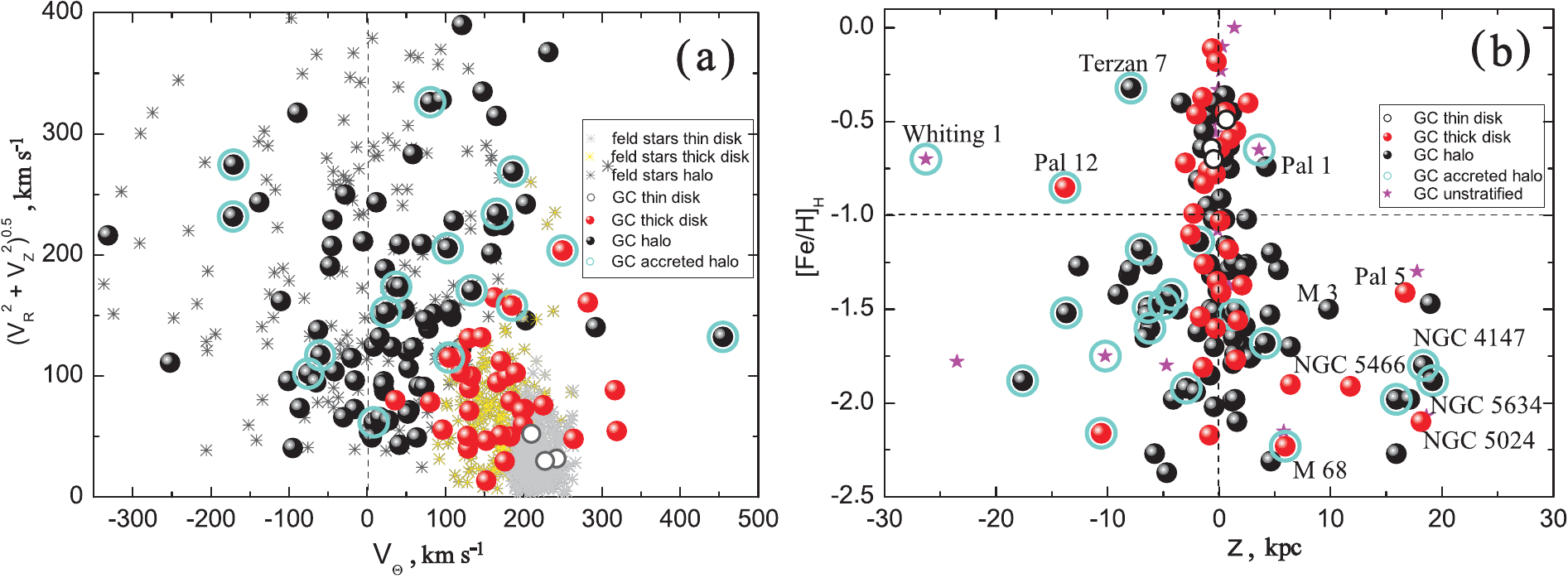}
\caption{The Tumre diagram for globular clusters and field 
         stars from \cite{12} (a) and the dependence of 
         metallicity on the distance to the galactic plane (b). 
         The field stars are indicated by: light gray 
         snowflakes-the thin disk, yellow-the thick disk, 
         dark gray-the halo. The large circles denoting 
         clusters belonging to the thin disk by kinematic 
         features are empty, those of the thick disk are red, 
         and those of the halo are dark gray. The stars 
         denote unstratified clusters. Circled are clusters 
         known as lost by dwarf galaxies. The values [Fe/H] 
         are taken from the catalog \cite{6}.}
\label{fig1}
\end{figure*}

\newpage

\begin{figure*}
\centering
\includegraphics[angle=0,width=0.99\textwidth,clip]{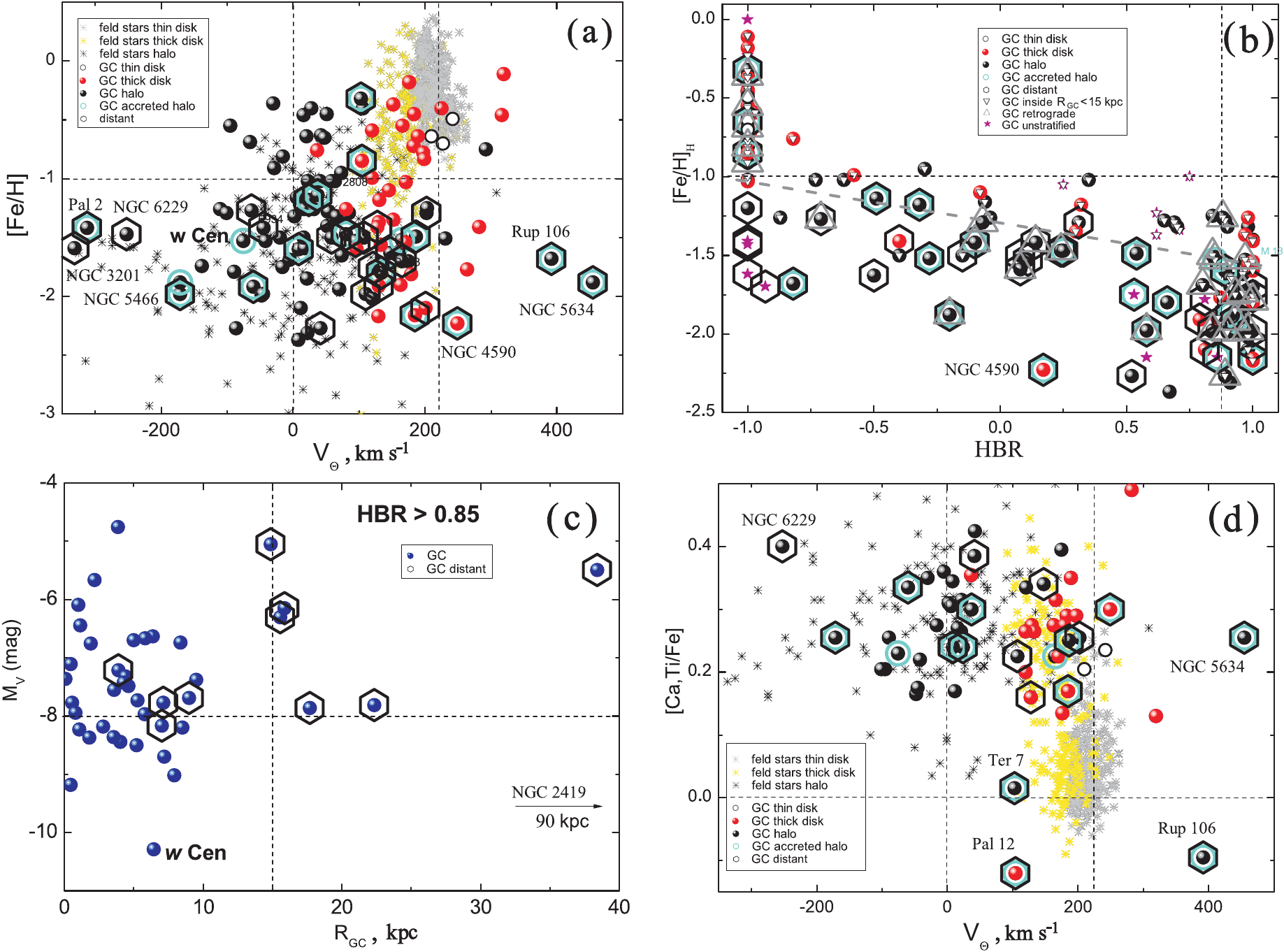}
\caption{The relation between ``spectroscopic metallicity'' a
         nd rotation speed around the galactic center for the 
         field stars and globular clusters (a); the relation 
         between metallicity from catalog \cite{6} and the 
         color of the horizontal branch clusters (b); the 
         relation between absolute magnitude and galactocentric 
         distance of clusters with extremely blue horizontal 
         branches (c) and the relation between relative 
         abundances of $\alpha$-elements and rotation speed 
         around the galactic center for field stars and 
         clusters (d). Large hexagons around circles are 
         distant clusters ($R_{GC}$ or $R_{max} > 15$~kpc); 
         light gray triangles around circles are clusters 
         in retrograde orbits; white triangles inside the 
         icons are clusters lying inside the solar circle 
         ($R_{G} < 8$~kpc). The dotted horizontal lines 
         are drawn through $\rm{[Fe/H]} = -1.0$ (a), (b) and 
         [$\alpha$/Fe] = 0.0 (d); the dotted vertical lines 
         are $V_{\Theta} = 0$ inside the solar circle 
         ($R_{G} < 8$~kpc). The dashed horizontal lines are 
         drawn through $\rm{[Fe/H]} = -1.0$ (a), (b) and 
         [$\alpha$/Fe] = 0.0 (d), and the vertical lines 
         are $V_{\Theta} = 0$ and 220~km\,s$^{-1}$ (a), (d) 
         and HBR = 0.85 (b); the inclined line is drawn 
         ``by eye'' and separates the positions of internal 
         and external clusters (b). Other designations are 
         the same as in Fig.~1. The names of clusters, 
         which deviate far from the average for the 
         respective subsystems, are inscribed.}
\label{fig2}
\end{figure*}

\newpage

\begin{figure*}
\centering
\includegraphics[angle=0,width=0.99\textwidth,clip]{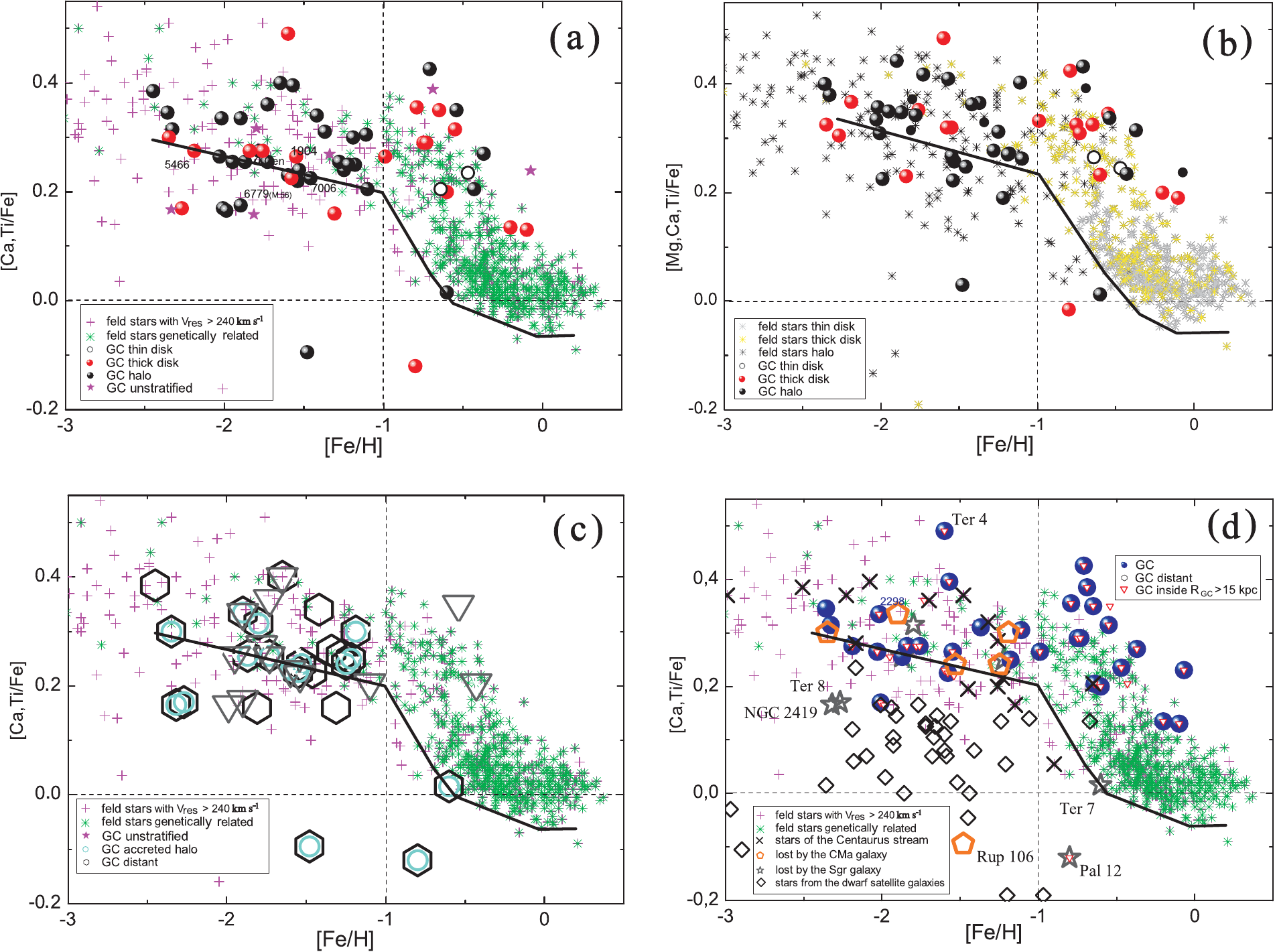}
\caption{The dependence of the relative abundances averaged over 
         two $\alpha$-elements (Ca and Ti) (a), (c), (d) and 
         four $\alpha$-elements (Mg, Ca, Si and Ti) (b) 
         on metallicity for the field stars from paper \cite{12} 
         and globular clusters (due to the lack of silicon 
         abundances in the field stars, they are averaged over 
         three elements). The stars of the field and clusters 
         of different subsystems are indicated as in 
         Fig.~1a (a), (b). Green snowflakes are genetically 
         related field stars with $V_{res} > 240$~km\,s$^{-1}$, 
         pink crosses are higher-velocity field stars (a) (c), (d); 
         outer and inner clusters in retrograde orbits are 
         indicated as in Figs.~2b and 2c. Large orange 
         pentagons are clusters lost by the CMA galaxy; 
         large gray stars are clusters lost by the Sgr 
         galaxy; diamonds are dwarf satellite galaxy stars 
         from papers \cite{35},\cite{37},\cite{38}, large 
         oblique crosses are Centauri stream stars (d). 
         Broken curve line is a drawn ``by eye'' lower envelope 
         for genetically related field stars (a)--(d).}
\label{fig3}
\end{figure*}

\newpage
\clearpage

\newpage
\begin{landscape}

\begin{table}[t!]

\caption{%
Parameters and labels of globular cluster belonging to subsystems and groups}
\bigskip
\begin{center}
\begin{tabular}{c|c|c|c|c|c|c|c|c|c|c|c|c|c|c|c}
\hline \hline

\multicolumn{1}{c|}{\parbox{1.6cm}{Name}}&
\multicolumn{1}{c|}{\parbox{0.9cm}{$x$, kpc}}&
\multicolumn{1}{c|}{\parbox{0.9cm}{$y$, kpc}}&
\multicolumn{1}{c|}{\parbox{0.9cm}{$z$, kpc}}&
\multicolumn{1}{c|}{\parbox{1.2cm}{$\rm{[Fe/H]}_{H}$}}&
\multicolumn{1}{c|}{\parbox{1.2cm}{$\rm{[Fe/H]}_{sp}$}}&
\multicolumn{1}{c|}{\parbox{1.5cm}{[CaTi/Fe]}}&
\multicolumn{1}{c|}{\parbox{2.5cm}{[MgSiCaTi/Fe]}}&
\multicolumn{1}{c|}{\parbox{0.8cm}{$V_{R}$, km\,s$^{-1}$}}&
\multicolumn{1}{c|}{\parbox{0.8cm}{$V_{\Theta}$, km\,s$^{-1}$}}&
\multicolumn{1}{c|}{\parbox{0.8cm}{$V_{Z}$, km\,s$^{-1}$}}&
\multicolumn{1}{c|}{\parbox{0.8cm}{$R_{GC}$, kpc}}&
\multicolumn{1}{c|}{\parbox{1.0cm}{$M_{V}$, mag}}&
\multicolumn{1}{c|}{\parbox{0.8cm}{HBR}}&
\multicolumn{1}{c|}{\parbox{0.6cm}{sub system}}&
\multicolumn{1}{c}{\parbox{0.8cm}{group}}\\

\hline
   1    &  2   & 3   &  4 & 5& 6& 7  &  8 & 9  & 10 & 11 & 12 & 13 & 14 &15&16\\
\hline
NGC 104&1.79& -2.47& -3.03& -0.72& -0.75& 0.29& 0.43& 12.74& 182.44& 48.35&
6.97& -9.42& -0.99& TD& i, t\\
NGC 288&-0.08& 0.04& -8.10& -1.32& -1.37& 0.31& 0.49& -9.81& 2.17& 52.37&
8.38& -6.74& 0.98& H& t\\
NGC 362&3.00& -4.89& -6.00& -1.26& -1.18& 0.25& 0.36& 111.21& 16.50& -100.71&
7.21& -8.41& -0.87& H& i, t\\
NGC 1261&0.09& -9.82& -12.63& -1.27& -& -& -& -115.17& -63.28& 76.33& 12.80&
-7.81& -0.71& H& o, r\\
NGC 1851&-4.30& -9.02& -7.00& -1.18& -1.25& 0.24& 0.42& 117.79& 23.78& -97.21&
15.49& -8.33& -0.32& H& o, a\\
NGC 1904& -7.46& -8.06& -6.18& -1.60& -1.53& 0.24& 0.34& 60.30& 8.17& 13.86&
17.70& -7.86& 0.89& H& o, a\\
NGC 2298& -4.20& -9.28& -2.92& -1.92& -1.90& 0.34& 0.59& -101.06& -60.01&
59.26& 15.57& -6.30& 0.93& H& o, r, a\\
NGC 2419& -74.70& -0.50& 35.20& -2.15& -2.32& 0.17& -& -& -& -& 89.90& -9.58&
0.86& -& o, a\\
NGC 2808& 1.93& -8.92& -1.81& -1.14& -1.19& 0.30& 0.36& -173.32& 36.61& 8.29&
10.96& -9.39& -0.49& H& o, a\\
NGC 3201& 0.63& -5.00& 0.77& -1.59& -1.49& -& -& -149.15& -331.61& 156.54&
9.15& -7.46& 0.08& H& o, r\\
NGC 4147& -1.23& -3.98& 18.33& -1.80& -& -& -& -29.86& 133.33& 168.17& 10.33&
-6.16& 0.66& H& a\\
NGC 4372& 2.49& -4.14& -0.84& -2.17& -2.19& 0.28& 0.49& 9.47& 128.78& 38.88&
7.14& -7.77& 1.00& TD& i, o, t\\
NGC 4590& 4.04& -7.10& 5.94& -2.23& -2.35& 0.30& 0.43& -203.47& 249.44& -7.17&
8.28& -7.35& 0.17& TD& o, a\\
NGC 4833& 3.23& -4.87& -0.82& -1.85& -2.03& 0.27& 0.45& 71.46& 20.96& -63.58&
7.02& -8.16& 0.93& H& i, t\\
NGC 5024& 2.91& -1.49& 18.11& -2.10& -& -& -& -19.56& 200.89& -70.16& 5.59&
-8.70& 0.81& TD& o\\
NGC 5053& 2.83& -1.28& 15.90& -2.27& -2.45& 0.39& -& -25.65& 41.15& 34.65&
5.62& -6.72& 0.52& H& o\\
NGC 5139& 3.11& -3.82& 1.32& -1.53& -1.60& 0.23& -& -63.14& -75.43& -80.04&
6.45& -10.29& 0.85& H& i, r, a\\
NGC 5272& 1.45& 1.32& 9.81& -1.50& -1.46& 0.23& 0.33& -63.77& 107.46& -134.72&
6.98& -8.93& 0.08& H& o\\
NGC 5286& 6.98& -7.86& 1.96& -1.69& -1.73& 0.36& 0.56& -206.77& -5.45& 43.78&
7.97& -8.61& 0.80& H& i, r\\
NGC 5466& 3.48& 3.15& 15.92& -1.98& -1.73& 0.26& 0.32& 60.37& -171.64& 223.68&
5.76& -6.96& 0.58& H& o, r, a\\
NGC 5634& 15.72& -5.04& 19.17& -1.88& -1.87& 0.26& 0.46& -101.18& 454.92&
85.01& 8.97& -7.69& 0.91& H& o, a, t\\
NGC 5694& 25.60& -14.15& 17.13& -1.98& -& -& -& -168.29& 110.46& -154.06&
22.35& -7.81& 1.00& H& o\\
NGC 5824& 25.74& -13.37& 11.76& -1.91& -& -& -& -130.05& 129.51& 17.33& 21.98&
-8.84& 0.79& TD& o\\
NGC 5897& 10.48& -3.22& 6.41& -1.90& -1.84& 0.28& 0.31& 132.78& 162.75& 97.81&
3.89& -7.21& 0.86& TD& i, t\\
NGC 5904& 4.99& 0.34& 5.32& -1.29& -1.28& 0.26& 0.37& -209.02& 202.43& -121.93&
3.33& -8.81& 0.31& H& i, o\\
NGC 5927& 6.16& -4.06& 0.63& -0.49& -0.47& 0.24& 0.33& -30.50& 242.14& 9.54&
4.59& -7.80& -1.00& D& i, t\\
NGC 5946& 10.36& -6.58& 0.90& -1.29& -& -& -& 52.88& 84.24& 141.70& 6.89&
-7.20& 0.69& H& i, t\\

\hline

\end{tabular}
\end{center}

\end{table}
\end{landscape}

\newpage
\begin{landscape}

\begin{table}[t!]


\begin{center}
\begin{tabular}{c|c|c|c|c|c|c|c|c|c|c|c|c|c|c|c|c}

\hline \hline

\multicolumn{1}{c|}{\parbox{1.6cm}{Name}}&
\multicolumn{1}{c|}{\parbox{0.9cm}{$x$, kpc}}&
\multicolumn{1}{c|}{\parbox{0.9cm}{$y$, kpc}}&
\multicolumn{1}{c|}{\parbox{0.9cm}{$z$, kpc}}&
\multicolumn{1}{c|}{\parbox{1.2cm}{$\rm{[Fe/H]}_{H}$}}&
\multicolumn{1}{c|}{\parbox{1.2cm}{$\rm{[Fe/H]}_{sp}$}}&
\multicolumn{1}{c|}{\parbox{1.5cm}{[CaTi/Fe]}}&
\multicolumn{1}{c|}{\parbox{2.5cm}{[MgSiCaTi/Fe]}}&
\multicolumn{1}{c|}{\parbox{0.8cm}{$V_{R}$, km\,s$^{-1}$}}&
\multicolumn{1}{c|}{\parbox{0.8cm}{$V_{\Theta}$, km\,s$^{-1}$}}&
\multicolumn{1}{c|}{\parbox{0.8cm}{$V_{Z}$, km\,s$^{-1}$}}&
\multicolumn{1}{c|}{\parbox{0.8cm}{$R_{GC}$, kpc}}&
\multicolumn{1}{c|}{\parbox{1.0cm}{$M_{V}$, mag}}&
\multicolumn{1}{c|}{\parbox{0.8cm}{HBR}}&
\multicolumn{1}{c|}{\parbox{0.6cm}{sub system}}&
\multicolumn{1}{c}{\parbox{0.8cm}{group}}\\

\hline
 
 1    &  2   & 3   &  4 & 5& 6& 7  &  8 & 9  & 10 & 11 & 12 & 13 & 14 &15&16\\

\hline

NGC 5986& 9.23& -3.91& 2.36& -1.59& -& -& -& 47.97& 17.56& -36.09& 4.02&
-8.44& 0.97& H& i, t\\
NGC 6093& 8.14& -1.05& 2.90& -1.75& -1.78& 0.28& 0.46& 6.50& -16.49& -72.35&
1.06& -8.23& 0.93& H& i, r\\
NGC 6101& 10.75& -9.77& -4.12& -1.98& -& -& -& -49.04& -374.20& -171.01&
10.07& -6.91& 0.84& H& o, r\\
NGC 6121& 2.09& -0.33& 0.61& -1.16& -1.11& 0.31& 0.54& -49.32& 5.71& -3.15&
6.22& -7.20& -0.06& H& i, t\\
NGC 6139& 9.93& -3.16& 1.27& -1.65& -& -& -& 43.25& 158.63& 196.63& 3.55&
-8.36& 0.91& H& i, t\\
NGC 6144& 9.63& -1.36& 2.73& -1.76& -& -& –& 363.16& 120.77& -141.76& 1.90&
-6.75& 1.00& H& i, o\\
NGC 6171& 5.79& 0.34& 2.46& -1.02& -1.05& -& -& 0.63& 63.51& -49.10& 2.53&
-7.13& -0.73& H& i, t\\
NGC 6205& 2.72& 4.54& 4.58& -1.53& -1.54& 0.22& 0.30& 64.24& -42.12& -81.63&
7.19& -8.70& 0.97& H& i, r\\
NGC 6218& 4.06& 1.14& 2.08& -1.37& -1.35& -& -& -12.21& 130.14& -89.04& 4.40&
-7.32& 0.97& TD& i, t\\
NGC 6229& 6.29& 21.44& 18.95& -1.47& -1.65& 0.40& -& 94.85& -252.56& 57.56&
21.53& -8.05& 0.24& H& o, r\\
NGC 6235& 9.43& -0.18& 2.27& -1.28& -& -& -& 106.52& 52.41& -6.39& 1.14&
-6.44& 0.89& H& i, t\\
NGC 6254& 3.82& 1.03& 1.69& -1.56& -1.55& 0.27& 0.43& -84.39& 132.81& 52.84&
4.60& -7.48& 0.98& TD& i, t\\
NGC 6256& 9.07& -1.96& 0.54& -1.02& -& -& -& -200.40& 40.69& 58.79& 2.11&
-6.52& -1.00& H& i, t\\
NGC 6266& 6.60& -0.74& 0.85& -1.18& -1.08& -& -& 30.09& 130.42& 64.43& 1.85&
-9.19& 0.32& TD& i, t\\
NGC 6273& 8.37& -0.46& 1.39& -1.74& -& -& -& -152.04& -138.65& 189.91& 0.46&
-9.18& 0.96& H& i, r\\
NGC 6284& 14.08& -0.41& 2.47& -1.26& -& -& -& 1.31& -110.79& 161.98& 5.79&
-7.97& 0.88& H& i, r\\
NGC 6287& 8.25& 0.02& 1.61& -2.10& -2.01& 0.17& 0.41& 231.08& 11.67& 75.95&
0.06& -7.36& 0.98& H& i, t\\
NGC 6293& 8.71& -0.36& 1.20& -1.99& -1.99& 0.17& 0.30& -144.69& -48.13& -124.02&
0.55& -7.77& 0.90& H& i, r\\
NGC 6304& 5.96& -0.43& 0.56& -0.45& -& -& -& 73.80& 183.05& 28.35& 2.38&
-7.32& -1.00& TD& i, t\\
NGC 6316& 11.43& -0.56& 1.15& -0.45& -& -& -& 67.47& 51.19& 99.83& 3.18&
-8.35& -1.00& H& i, t\\
NGC 6333& 8.12& 0.79& 1.54& -1.77& -& -& -& 13.66& 263.62& 46.02& 0.81& -7.94&
0.87& TD& i, t\\
NGC 6341& 2.45& 6.18& 4.63& -2.31& -2.33& 0.32& 0.51& 57.63& 21.80& 65.87&
8.51& -8.20& 0.91& H& t\\
NGC 6342& 8.94& 0.77& 1.54& -0.55& -0.55& 0.32& 0.46& 17.35& 165.92& -93.08&
1.00& -6.44& -1.00& TD& i, t\\
NGC 6352& 5.27& -1.77& -0.70& -0.64& -0.64& 0.21& 0.35& 52.30& 209.39& -1.01&
3.51& -6.48& -1.00& D& i, t\\
NGC 6355& 9.20& -0.10& 0.90& -1.37& -& -& -& -& -& -& 1.40& -8.08& 0.62&
-& i, t\\
NGC 6356& 14.27& 1.68& 2.59& -0.40& -& -& -& 19.37& 224.75& 73.24& 6.20&
-8.52& -1.00& TD& i, t\\
NGC 6362& 5.90& -4.04& -2.26& -0.99& -0.99& 0.27& 0.44& 35.37& 119.36& 96.72&
4.71& -6.94& -0.58& TD& i, t\\

\hline

\end{tabular}
\end{center}

\end{table}
\end{landscape}

\newpage

\begin{landscape}

\begin{table}[t!]

\begin{center}
\begin{tabular}{c|c|c|c|c|c|c|c|c|c|c|c|c|c|c|c|c}

\hline \hline

\multicolumn{1}{c|}{\parbox{1.6cm}{Name}}&
\multicolumn{1}{c|}{\parbox{0.9cm}{$x$, kpc}}&
\multicolumn{1}{c|}{\parbox{0.9cm}{$y$, kpc}}&
\multicolumn{1}{c|}{\parbox{0.9cm}{$z$, kpc}}&
\multicolumn{1}{c|}{\parbox{1.2cm}{$\rm{[Fe/H]}_{H}$}}&
\multicolumn{1}{c|}{\parbox{1.2cm}{$\rm{[Fe/H]}_{sp}$}}&
\multicolumn{1}{c|}{\parbox{1.5cm}{[CaTi/Fe]}}&
\multicolumn{1}{c|}{\parbox{2.5cm}{[MgSiCaTi/Fe]}}&
\multicolumn{1}{c|}{\parbox{0.8cm}{$V_{R}$, km\,s$^{-1}$}}&
\multicolumn{1}{c|}{\parbox{0.8cm}{$V_{\Theta}$, km\,s$^{-1}$}}&
\multicolumn{1}{c|}{\parbox{0.8cm}{$V_{Z}$, km\,s$^{-1}$}}&
\multicolumn{1}{c|}{\parbox{0.8cm}{$R_{GC}$, kpc}}&
\multicolumn{1}{c|}{\parbox{1.0cm}{$M_{V}$, mag}}&
\multicolumn{1}{c|}{\parbox{0.8cm}{HBR}}&
\multicolumn{1}{c|}{\parbox{0.6cm}{sub system}}&
\multicolumn{1}{c}{\parbox{0.8cm}{group}}\\

\hline
 1  &  2   & 3   &  4 & 5& 6& 7  &  8 & 9  & 10 & 11 & 12 & 13 & 14 &15&16\\

\hline

NGC 6366& 3.28& 1.09& 0.99& -0.59& -0.60& 0.20& 0.31& 91.01& 119.87& -71.95&
5.13& -5.77& -0.97& TD& i, t\\
NGC 6380& 10.70& -1.90& -0.60& -0.75& -& -& -& -& -& -& 3.30& -7.46& -1.00&
-& i, t\\
NGC 6388& 11.06& -2.85& -1.35& -0.55& -0.43& 0.21& 0.31& -28.39& -95.50&
-28.79& 3.97& -9.42& -1.00& H& i, r\\
NGC 6397& 2.00& -0.80& -0.46& -2.02& -2.02& 0.34& 0.48& 35.85& 120.22& -118.07&
6.35& -6.63& 0.98& H& i, t\\
NGC 6401& 7.47& 0.45& 0.52& -1.02& -& -& -& 90.11& 58.35& 84.46& 0.95& -7.90&
0.35& H& i, t\\
NGC 6402& 7.84& 3.06& 2.22& -1.28& -& -& -& -69.40& 53.76& -17.47& 3.09&
-9.12& 0.65& H& i, t\\
NGC 6426& 17.40& 9.30& 5.80& -2.15& -& -& -& -& -& -& 14.40& -6.69& 0.58&
-& t\\
NGC 6440& 7.91& 1.07& 0.53& -0.36& -0.54& 0.35& 0.45& 41.03& -30.69& 51.44&
1.14& -8.75& -1.00& H& i, r\\
NGC 6441& 9.60& -1.09& -0.85& -0.46& -0.37& 0.27& 0.42& 17.12& 16.01& 63.81&
1.70& -9.64& -1.00& H& i, t\\
NGC 6453& 10.84& -0.81& -0.74& -1.50& -& -& -& -103.73& -20.36& -49.58& 2.67&
-6.88& 0.84& H& i, r\\
NGC 6496& 11.18& -2.37& -2.02& -0.46& -& -& -& -28.90& 316.36& -83.41& 3.73&
-7.23& -1.00& TD& i, t\\
NGC 6517& 10.00& 3.50& 1.30& -1.23& -& -& -& -& -& -& 4.20& -8.28& 0.62&
-& i, t\\
NGC 6522& 7.70& 0.10& -0.50& -1.34& -& -& -& -& -& -& 0.60& -7.67& 0.71&
-& i, t\\
NGC 6528& 7.90& 0.20& -0.60& -0.11& -0.10& 0.13& 0.25& 53.85& 319.02& 8.55&
0.60& -6.56& -1.00& TD& i, t\\
NGC 6535& 5.95& 3.05& 1.23& -1.79& -1.95& 0.26& 0.47& 245.92& -89.78& 200.80&
3.85& -4.75& 1.00& H& i, r\\
NGC 6539& 7.33& 2.79& 0.93& -0.63& -0.71& 0.43& 0.58& -130.55& 42.16& 114.61&
2.95& -8.30& -1.00& H& i, t\\
NGC 6540& 3.49& 0.20& -0.20& -1.35& -& -& -& 8.42& 151.86& 10.61& 4.82& -5.38&
0.30& TD& i, t\\
NGC 6541& 7.13& -1.35& -1.44& -1.81& -1.76& 0.28& 0.47& 95.40& 178.63& -28.83&
1.78& -8.37& 1.00& TD& i, t\\
NGC 6544& 2.49& 0.25& -0.10& -1.40& -& -& -& 2.65& 51.43& -69.10& 5.82& -6.66&
1.00& H& i, t\\
NGC 6553& 4.67& 0.43& -0.25& -0.18& -0.20& 0.14& 0.27& 9.61& 175.77& 27.95&
3.65& -7.77& -1.00& TD& i, t\\
NGC 6558& 6.36& 0.02& -0.67& -1.32& -& -& -& 137.66& 78.12& -19.88& 1.94&
-6.46& 0.70& H& i, t\\
NGC 6569& 8.44& 0.07& -0.99& -0.76& -0.79& 0.36& 0.57& -45.09& 35.89& -66.22&
0.16& -8.30& -0.82& TD& i, t\\
NGC 6584& 11.87& -3.82& -3.67& -1.50& -& -& -& 136.53& 58.05& -248.00& 5.23&
-7.68& -0.15& H& i, o\\
NGC 6624& 7.80& 0.40& -1.10& -0.44& -0.69& 0.39& 0.52& -& -& -& 1.20& -7.49&
-1.00& -& i, t\\
NGC 6626& 5.62& 0.77& -0.55& -1.32& -& -& -& -27.07& 65.64& -87.47& 2.79&
-8.18& 0.90& H& i, t\\
NGC 6637& 8.06& 0.24& -1.46& -0.64& -& -& -& -76.22& 20.18& 53.10& 0.34&
-7.64& -1.00& H& i, t\\

\hline

\end{tabular}
\end{center}

\end{table}
\end{landscape}

\newpage

\begin{landscape}

\begin{table}[t!]

\begin{center}
\begin{tabular}{c|c|c|c|c|c|c|c|c|c|c|c|c|c|c|c|c}

\hline \hline

\multicolumn{1}{c|}{\parbox{1.6cm}{Name}}&
\multicolumn{1}{c|}{\parbox{0.9cm}{$x$, kpc}}&
\multicolumn{1}{c|}{\parbox{0.9cm}{$y$, kpc}}&
\multicolumn{1}{c|}{\parbox{0.9cm}{$z$, kpc}}&
\multicolumn{1}{c|}{\parbox{1.2cm}{$\rm{[Fe/H]}_{H}$}}&
\multicolumn{1}{c|}{\parbox{1.2cm}{$\rm{[Fe/H]}_{sp}$}}&
\multicolumn{1}{c|}{\parbox{1.5cm}{[CaTi/Fe]}}&
\multicolumn{1}{c|}{\parbox{2.5cm}{[MgSiCaTi/Fe]}}&
\multicolumn{1}{c|}{\parbox{0.8cm}{$V_{R}$, km\,s$^{-1}$}}&
\multicolumn{1}{c|}{\parbox{0.8cm}{$V_{\Theta}$, km\,s$^{-1}$}}&
\multicolumn{1}{c|}{\parbox{0.8cm}{$V_{Z}$, km\,s$^{-1}$}}&
\multicolumn{1}{c|}{\parbox{0.8cm}{$R_{GC}$, kpc}}&
\multicolumn{1}{c|}{\parbox{1.0cm}{$M_{V}$, mag}}&
\multicolumn{1}{c|}{\parbox{0.8cm}{HBR}}&
\multicolumn{1}{c|}{\parbox{0.6cm}{sub system}}&
\multicolumn{1}{c}{\parbox{0.8cm}{group}}\\

\hline

1  &  2   & 3   &  4 & 5& 6& 7  &  8 & 9  & 10 & 11 & 12 & 13 & 14 &15&16\\

\hline

NGC 6638& 8.06& 1.12& -1.02& -0.95& -& -& -& 81.56& 72.57& 40.62& 1.14& -7.13&
-0.30& H& i, t\\
NGC 6642& 7.44& 1.29& -0.85& -1.26& -& -& -& 104.54& 31.04& -66.93& 1.55&
-6.77& -0.04& H& i, t\\
NGC 6652& 9.21& 0.25& -1.85& -0.81& -& -& -& -91.84& -16.04& 28.43& 0.94&
-6.68& -1.00& H& i, r\\
NGC 6656& 3.13& 0.54& -0.42& -1.70& -1.57& 0.40& 0.55& 172.23& 174.22& -143.79&
5.20& -8.50& 0.91& H& i, t\\
NGC 6681& 8.48& 0.42& -1.88& -1.62& -& -& -& 137.00& 161.44& -176.82& 0.46&
-7.11& 0.96& H& i, t\\
NGC 6712& 6.04& 2.86& -0.50& -1.02& -& -& -& 118.47& 22.38& -146.65& 3.65&
-7.50& -0.62& H& i, t\\
NGC 6715& 25.29& 2.48& -6.38& -1.49& -1.22& 0.25& 0.25& 211.70& 186.21& 165.82&
17.17& -10.01& 0.54& H& i, o, a, t\\
NGC 6717& 6.80& 1.55& -1.34& -1.26& -& -& -& -55.34& 80.35& 55.64& 2.16&
-5.66& 0.98& TD& i, t\\
NGC 6723& 8.21& 0.01& -2.56& -1.10& -& -& -& 123.07& 145.44& -48.05& 0.09&
-7.84& -0.08& TD& i, t\\
NGC6749& 6.21& 4.54& -0.30& -1.60& -& -& -& -0.25& 96.05& 55.77& 5.00& -6.70&
1.00& TD& i, t\\
NGC 6752& 3.22& -1.40& -1.69& -1.54& -1.58& 0.23& 0.43& -14.00& 170.39& 49.27&
5.27& -7.73& 1.00& TD& i, t\\
NGC 6760& 5.88& 4.29& -0.50& -0.40& -& -& -& 143.62& 105.55& -57.55& 4.93&
-7.86& -1.00& H& i, t\\
NGC 6779& 4.50& 8.70& 1.44& -1.98& -1.90& 0.18& -& 166.86& -45.45& 122.97&
9.50& -7.38& 0.98& H& i, r\\
NGC 6809& 4.81& 0.74& -2.09& -1.94& -1.93& -& -& -204.06& 69.94& -45.00&
3.57& -7.55& 0.87& H& i, t\\
NGC 6838& 2.08& 3.17& -0.30& -0.78& -0.73& 0.29& 0.41& 15.94& 196.98& 69.77&
6.98& -5.60& -1.00& TD& i, t\\
NGC 6864& 15.54& 5.75& -7.99& -1.29& -1.10& 0.21& 0.35& -95.59& -101.44&
-8.19& 9.25& -8.55& -0.07& H& r\\
NGC 6934& 8.83& 11.35& -4.92& -1.47& -& -& -& -320.11& 80.44& 61.31& 11.36&
-7.46& 0.25& H& o, a\\
NGC 6981& 11.56& 8.14& -9.07& -1.42& -& -& -& -162.16& -44.68& 161.76& 8.77&
-7.04& 0.14& H& o, r\\
NGC 7006& 17.20& 34.80& -13.70& -1.52& -1.55& 0.23& 0.36& -180.81& 165.32&
148.41& 38.50& -7.68& -0.28& H& o, a\\
NGC 7078& 3.83& 8.21& -4.68& -2.37& -2.36& 0.35& 0.53& 102.63& 7.87& -70.09&
9.35& -9.17& 0.67& H& t\\
NGC 7089& 5.52& 7.42& -6.66& -1.65& -& -& -& 102.76& 73.79& -103.95& 7.93&
-9.02& 0.92& H& i, t\\
NGC 7099& 4.81& 2.47& -5.76& -2.27& -2.31& -& -& -0.43& -86.39& 73.72& 4.28&
-7.43& 0.89& H& i, r\\
NGC 7492& 7.00& 9.40& -23.50& -1.78& -1.81& 0.16& 0.42& -& -& -& 25.30& -5.77&
0.81& -& o\\
1636-283& 7.36& -1.05& 1.59& -1.50& -& -& -& 35.09& 15.86& 127.32& 1.41&
-3.97& -0.40& H& i, t\\
2MS-GC01& 3.50& 0.70& 0.00& -& -& -& -& -& -& -& 4.50& -& -& -& i, t\\
2MS-GC02& 4.90& 0.80& -0.10& -1.08& -& -& -& -& -& -& 3.20& -& -&-& i, t\\

\hline

\end{tabular}
\end{center}

\end{table}
\end{landscape}

\newpage

\begin{landscape}

\begin{table}[t!]

\begin{center}
\begin{tabular}{c|c|c|c|c|c|c|c|c|c|c|c|c|c|c|c|c}

\hline \hline

\multicolumn{1}{c|}{\parbox{1.6cm}{Name}}&
\multicolumn{1}{c|}{\parbox{0.9cm}{$x$, kpc}}&
\multicolumn{1}{c|}{\parbox{0.9cm}{$y$, kpc}}&
\multicolumn{1}{c|}{\parbox{0.9cm}{$z$, kpc}}&
\multicolumn{1}{c|}{\parbox{1.2cm}{$\rm{[Fe/H]}_{H}$}}&
\multicolumn{1}{c|}{\parbox{1.2cm}{$\rm{[Fe/H]}_{sp}$}}&
\multicolumn{1}{c|}{\parbox{1.5cm}{[CaTi/Fe]}}&
\multicolumn{1}{c|}{\parbox{2.5cm}{[MgSiCaTi/Fe]}}&
\multicolumn{1}{c|}{\parbox{0.8cm}{$V_{R}$, km\,s$^{-1}$}}&
\multicolumn{1}{c|}{\parbox{0.8cm}{$V_{\Theta}$, km\,s$^{-1}$}}&
\multicolumn{1}{c|}{\parbox{0.8cm}{$V_{Z}$, km\,s$^{-1}$}}&
\multicolumn{1}{c|}{\parbox{0.8cm}{$R_{GC}$, kpc}}&
\multicolumn{1}{c|}{\parbox{1.0cm}{$M_{V}$, mag}}&
\multicolumn{1}{c|}{\parbox{0.8cm}{HBR}}&
\multicolumn{1}{c|}{\parbox{0.6cm}{sub system}}&
\multicolumn{1}{c}{\parbox{0.8cm}{group}}\\

\hline

1  &  2   & 3   &  4 & 5& 6& 7  &  8 & 9  & 10 & 11 & 12 & 13 & 14 &15&16\\

\hline

AM1& -16.50& -80.10& -92.30& -1.70& -& -& -& -& -& -& 124.60& -4.71& -0.93&
-&o\\
AM4& 20.70& -17.20& 17.80& -1.30& -& -& -& -& -& -& 27.80& -1.60& -&-& o\\
Arp 2& 26.40& 4.00& -10.20& -1.75& -1.80& 0.32& 0.50& -& -& -& 21.40& -5.29&
0.53&-& o\\
BH 176& 16.10& -9.90& 1.40& 0.00& -& -& -& -& -& -& 12.90& -4.35& -1.00&-&
t\\
BH 261& 6.50& 0.40& -0.60& -1.30& -& -& -& -& -& -& 1.70& -& -&-& i, t\\
Djorg 1& 9.18& -0.53& -0.40& -1.51& -& -& -& -281.58& 231.15& 236.16& 1.03&
-6.26& -& H& i, t\\
Djorg 2& 13.77& 0.67& -0.60& -0.65& -& -& -& -128.51& 47.23& 87.17& 5.51&
-6.98& -1.00& H& i, t\\
E3& 1.50& -3.67& -1.37& -0.83& -& -& -& -16.74& 198.76& -57.46& 7.73& -2.77&
-& TD& i, t\\
Eridanus& -53.20& -41.70& -59.50& -1.43& -& -& -& -& -& -& 95.00& -5.14&
-1.00&& o\\
ESO-SC06& 20.40& -4.70& -4.70& -1.80& -& -& -& -& -& -& 14.00& -& -&-& t\\
FSR 1735& 9.10& -3.50& -0.30& -& -& -& -& -& -& -& 3.70& -& -&-& i, t\\
GLIMPSE01& 3.60& 2.20& 0.00& -& -& -& -& -& -& -& 4.90& -& -&-& i, t\\
GLIMPSE02& 5.40& 1.40& -0.10& -0.33& -& -& -& -& -& -& 3.00& -& -&-& i, t\\
HP 1& 8.20& -0.40& 0.30& -1.00& -& -& -& -& -& -& 0.50& -6.44& 0.75&-& i,
t\\
IC 1257& 22.67& 6.73& 6.40& -1.70& -& -& -& 125.78& 165.04& 288.49& 15.87&
-6.15& 1.00& H& o,\\
IC 1276& 8.59& 3.44& 0.92& -0.75& -& -& -& -37.49& 292.05& 135.07& 3.45&
-6.67& -1.00& H& i, t\\
IC 4499& 10.46& -13.70& -6.43& -1.53& -& -& -& -229.21& 94.99& -234.48& 13.87&
-7.33& 0.11& H& o\\
Ko 1& -2.50& -15.70& 45.60& -& -& -& -& -& -& -& 49.30& -& -&-& o\\
Ko 2& -30.20& -8.20& 15.00& -& -& -& -& -& -& -& 41.90& -& -&-& o\\
Liller 1& 8.20& -0.70& 0.00& -0.33& -& -& -& -& -& -& 0.80& -7.63& -1.00&-&
i, t\\
Lynga 7& 6.80& -4.10& -0.40& -1.01& -& -& -& -& -& -& 4.30& –& -1.00&-& i,
t\\
Pal 1& -6.80& 8.10& 3.60& -0.65& -& -& -& -& -& -& 17.20& -2.47& -1.00&-&
o, a\\
Pal 10& 3.60& 4.70& 0.30& -0.10& -& -& -& -& -& -& 6.40& -5.79& -1.00&-&
i, t\\
Pal 11& 10.31& 6.40& -3.38& -0.40& -& -& -& 63.32& 27.34& -1.15& 6.71& -6.86&
-1.00& H& i, t\\
Pal 12& 10.85& 6.39& -13.83& -0.85& -0.80& -0.12& -0.02& 114.64& 104.40&
-11.88& 6.88& -4.48& -1.00& TD& i, o, a\\
Pal 13& 1.00& 19.10& -17.60& -1.88& -& -& -& 255.33& -171.54& -100.45& 26.90&
-3.74& -0.20& H& o, r, a\\
Pal 14& 49.70& 27.30& 51.40& -1.62& -1.34& 0.27& 0.44& -& -& -& 71.60& -4.73&
-1.00&-& o\\
Pal 15& 38.90& 13.30& 18.60& -2.07& -& -& -& -& -& -& 38.40& -5.49& 1.00&&
o\\

\hline

\end{tabular}
\end{center}

\end{table}
\end{landscape}

\newpage

\begin{landscape}

\begin{table}[t!]

\begin{center}
\begin{tabular}{c|c|c|c|c|c|c|c|c|c|c|c|c|c|c|c|c}

\hline \hline

\multicolumn{1}{c|}{\parbox{1.6cm}{Name}}&
\multicolumn{1}{c|}{\parbox{0.9cm}{$x$, kpc}}&
\multicolumn{1}{c|}{\parbox{0.9cm}{$y$, kpc}}&
\multicolumn{1}{c|}{\parbox{0.9cm}{$z$, kpc}}&
\multicolumn{1}{c|}{\parbox{1.2cm}{$\rm{[Fe/H]}_{H}$}}&
\multicolumn{1}{c|}{\parbox{1.2cm}{$\rm{[Fe/H]}_{sp}$}}&
\multicolumn{1}{c|}{\parbox{1.5cm}{[CaTi/Fe]}}&
\multicolumn{1}{c|}{\parbox{2.5cm}{[MgSiCaTi/Fe]}}&
\multicolumn{1}{c|}{\parbox{0.8cm}{$V_{R}$, km\,s$^{-1}$}}&
\multicolumn{1}{c|}{\parbox{0.8cm}{$V_{\Theta}$, km\,s$^{-1}$}}&
\multicolumn{1}{c|}{\parbox{0.8cm}{$V_{Z}$, km\,s$^{-1}$}}&
\multicolumn{1}{c|}{\parbox{0.8cm}{$R_{GC}$, kpc}}&
\multicolumn{1}{c|}{\parbox{1.0cm}{$M_{V}$, mag}}&
\multicolumn{1}{c|}{\parbox{0.8cm}{HBR}}&
\multicolumn{1}{c|}{\parbox{0.6cm}{sub system}}&
\multicolumn{1}{c}{\parbox{0.8cm}{group}}\\

\hline

1  &  2   & 3   &  4 & 5& 6& 7  &  8 & 9  & 10 & 11 & 12 & 13 & 14 &15&16\\

\hline

Pal 2& -26.20& 4.37& -4.24& -1.42& -& -& -& 26.47& -312.64& 746.70& 34.78&
-8.01& -0.10& H& o, r, a\\
Pal 3& -34.30& -59.70& 61.70& -1.63& -1.42& 0.34& 0.48& -262.14& 147.14&
208.07& 95.70& -5.70& -0.50& H& o\\
Pal 4& -31.40& -12.90& 103.20& -1.41& -& -& -& -& -& -& 111.20& -6.02& -1.00&&
o\\
Pal 5& 16.20& 0.20& 16.70& -1.41& -1.31& 0.16& -& -48.01& 127.88& -15.66&
18.60& -5.17& -0.40& TD& o\\
Pal 6& 6.69& 0.25& 0.21& -0.91& -& -& -& -226.31& -28.69& 106.27& 1.63& -6.81&
-1.00& H& i, r\\
Pal 8& 11.94& 3.00& -1.47& -0.37& -& -& -& -40.69& 152.12& 22.66& 4.72& -5.52&
-1.00& TD& i, t\\
Pyxis& -5.77& -37.78& 4.69& -1.20& -& -& -& -83.14& 682.67& 82.20& 40.31&
-5.75& -1.00& H& o\\
Rup 106& 10.35& -17.31& 4.17& -1.68& -1.48& -0.10& 0.04& 6.09& 391.65& -464.16&
17.43& -6.35& -0.82& H& o, a\\
Terzan 1& 6.49& -0.28& 0.11& -1.03& -& -& -& -70.36& 170.75& -87.41& 1.83&
-4.90& -1.00& TD& i, t\\
Terzan 10& 5.80& 0.50& -0.20& -1.00& -& -& -& -& -& -& 2.30& -6.31& -1.00&-&
i, t\\
Terzan 12& 4.70& 0.70& -0.20& -0.50& -& -& -& -& -& -& 3.40& -4.14& -1.00&-&
i, t\\
Terzan 2& 9.47& -0.61& 0.38& -0.69& -& -& -& 82.25& -65.32& 47.20& 1.32&
-5.27& -1.00& H& i, r\\
Terzan 3& 25.18& -6.71& 4.22& -0.74& –& –& –& -176.64& 513.67& 277.94& 18.17&
-4.61& -1.00& H&-\\
Terzan 4& 7.20& -0.50& 0.20& -1.41& -1.60& 0.49& 0.65& -135.13& 281.86& -87.10&
1.00& -6.09& 1.00& TD& i, t\\
Terzan 5& 6.90& 0.50& 0.20& -0.23& -0.07& 0.23& 0.32& –& –& –& 1.20& -7.87&
-1.00& -& i, t\\
Terzan 6& 6.80& -0.20& -0.30& -0.56& –& –& –& –& –& –& 1.30& -7.67& -1.00&
-& i, t\\
Terzan 7& 21.57& 1.28& -7.89& -0.32& -0.60& 0.02& 0.02& 204.03& 102.80& 24.39&
13.33& -5.05& -1.00& H& o, a\\
Terzan 8& 22.99& 2.32& -10.56& -2.16& -2.27& 0.17& 0.40& 157.28& 184.87&
-15.49& 14.87& -5.05& 1.00& TD& o, a\\
Terzan 9& 7.10& 0.40& -0.20& -1.05& –& –& –& –& –& –& 1.10& -3.85& 0.25&
-& i, t\\
Ton 2& 7.78& -1.26& -0.47& -0.70& -& -& -& -24.05& 227.11& -17.70& 1.36&
-6.14& -1.00& D& i, t\\
UKS1& 7.47& 0.67& 0.10& -0.64& -0.65& 0.35& 0.43& 60.48& 189.39& 83.13& 1.07&
-6.88& -1.00& TD& i, t\\
Whiting 1& -13.90& 4.70& -26.30& -0.70& –& –& –& –& –& –& 34.50& –& –& –&
o, a\\

\hline

\end{tabular}
\end{center}

\end{table}

Belonging of clusters to galactic subsystems and groups:

T, TD and H -- thin disk, thick disk and halo, respectively;

i -- internal: located at a distance (or apogalactic radii of their 
orbits) less than 8~kpc;

o -- distant: located at a distance or have radii of orbits 
greater than 15~kpc;

r -- retrograde: the azimuthal components of their velocity are 
less than zero;

a -- accreted: the works of other authors prove their extragalactic origin;

t -- genetically related: the clusters not falling into any of the last 
three groups.

\end{landscape}



\begin{thebibliography}{}

\bibitem[1]{1} V. A. Marsakov, V. V. Koval' and M. L. Gozha, 
Astron. Rep. 63, 274 (2019). 
\bibitem[2]{2} T. V. Borkova and V. A. Marsakov, Astron. Rep. 
{\bf44}, 665 (2000).
\bibitem[3]{3} V. A. Marsakov and A. A. Suchkov, Astron. Rep. 
{\bf21}, 700 (1977).
\bibitem[4]{4} E. Carretta, IAU Symp. 317, 2015. (Eds. A. Bragaglia, 
M. Arnaboldi, M. Rejkuba \& D. Romano) 97 (2016).
\bibitem[5]{5} B. J. Pritzl, K. A. Venn and M. Irwin, Astron. J. 
{\bf130}, 2140 (2005).
\bibitem[6]{6} W. E. Harris, Astron. J. {\bf112}, 1487 (1996); 
2010 edition [arXiv:1012.3224].
\bibitem[7]{7} G. M. Eadie and W. E. Harris, Astrophys. J. {\bf829}, 
108 (2016). 
\bibitem[8]{8} A. A. Chemel, E. V. Glushkova, A. K. Dambis, 
A. S. Rastorguev, et al., Astrophys. Bull. {\bf73}, 162 (2018).
\bibitem[9]{9} A. A. Chemel, (private communication).
\bibitem[10]{10} D. A. VandenBerg, Astrophys. J. Sup, {\bf129}, 
315 (2000). 
\bibitem[11]{11} M. Salaris and A. Weiss, Astron. Astrophys. 
{\bf388}, 492 (2002).
\bibitem[12]{12} K. A. Venn, M. Irwin, M. D. Shetrone, et al., 
Astron. J, {\bf128}, 1177 (2004).
\bibitem[13]{13} T. Bensby, S. Feldsing and I. Lundstrem, Astron. 
Astrophys. {\bf410}, 527 (2003). 
\bibitem[14]{14} O. J.Eggen, D. Linden-Bell and A. Sandage, 
Astrophys. J. {\bf136}, 748 (1962).   
\bibitem[15]{15} M. G. Abadi, J. F. Navarro and M. Steinmetz, 
Monthly Not. Roy. Astron. Soc. {\bf365}, 747 (2006).   
\bibitem[16]{16} Y.-W. Lee, H. B. Gim and D. I. Casetti- Dinescu, 
Astrophys. J. {\bf661}, L49 (2007). 
\bibitem[17]{17} R. G. Gratton, E. Carretta and A. Bragaglia, 
Astron. Astrophys. Rev., {\bf20}, 50 (2012). 
\bibitem[18]{18} T. Decressin, G. Meynet, C. Charbonnel, et al., 
Astron. Astrophys. {\bf 464}, 1029 (2007). 
\bibitem[19]{19} P. Ventura and F. D'Antona, Astron. Astrophys. 
{\bf499}, 835 (2009).
\bibitem[20]{20} S. Jang, Y.-W. Lee, S.-J. Joo and C. Na, 
Monthly Not. Roy. Astron. Soc. {\bf443}, 15 (2014). 
\bibitem[21]{21} T. V. Borkova and V. A. Marsakov, 
Bull. Spec. Astrophys. Obs. {\bf54}, 61(2002).
\bibitem[22]{22} Y.-W. Lee, P. Demarque and R. Zinn, 
Astrophys. J. {\bf423}, 248 (1994). 
\bibitem[23]{23} V. A. Marsakov and T. V. Borkova, 
Astronomy. Lett. {\bf32}, 545 (2006).
\bibitem[24]{24} P. E. Nissen and W. J. Schuster, 
Astron. Astrophys. {\bf511}, L10 (2010). 
\bibitem[25]{25} V. V. Bobylev and A. T. Bajkova, 
Astron. Rep. {\bf61}, 551 (2017).
\bibitem[26]{26} D. A. Forbes and T. Bridges, 
Monthly Not. Roy. Astron. Soc. {\bf404}, 1203 (2010). 
\bibitem[27]{27} A. Mucciarelli, M. Bellazzini, R. Ibata, et al., 
Astron. Astrophys. {\bf605}, 46 (2017).
\bibitem[28]{28} S. L. J. Gibbons, V. Belokurov, and N. W. Evans, 
Monthly Not. Roy. Astron. Soc. {\bf464}, 794 (2017).
\bibitem[29]{29} V. Marsakov, T. Borkova, V. Koval' In: 
``Variable stars, the Galactic halo end Galaxy formation'', 
B.V.Kurarkin, Centenary Conf. Zvenigorod, Russia, 2009. 
(Eds. C. Sterken, N. Samus, L. Szabodos), Moscow: 
University Press, 133 (2010).
\bibitem[30]{30} T. Tshuchiya, D. Dinescu and V.I. Korchagin, 
Astrophys. J. Lett. {\bf589}, L29 (2003).
\bibitem[31]{31} M. G. Abadi, M. G.Navarro, M. Steinmetzand, 
and V. R. Eke, Astrophys. J. {\bf591}, 499 (2003).
\bibitem[32]{32} A. Meza, J. F. Navarro, M. G. Abadi and M. Steinmetz, 
Monthly Not. Roy. Astron. Soc. {\bf359}, 93 (2005).
\bibitem[33]{33} T. V. Borkova and V. A. Marsakov, 
Astron. Rep. {\bf49}, 405 (2005).
\bibitem[34]{34} V. A. Marsakov, M. L. Gozha, and V. V. Koval', 
Astron. Rep. {\bf62}, 50 (2018).
\bibitem[35]{35} J. T. Mackereth, R. P. Schiavon, J. Pfeffer, et al. 
Monthly Not. Roy. Astron. Soc. {\bf482}, 3426 (2018).
\bibitem[36]{36} V. A. Marsakov, M. L. Gozha, and V. V. Koval', 
Astron. Rep. {\bf63}, 203 в печати (2019).
\bibitem[37]{37} M. D. Shetrone, P. Cote, and W. L. W. Sargent, 
Astrophys. J. {\bf548}, 592 (2001).
\bibitem[38]{38} M. Shetrone, K. A. Venn, E. Tolstoy, et al., 
Astron. J. {\bf125}, 684 (2003).
\bibitem[39]{39} D. Geisler, V. V. Smith, G. Wallerstein, et al., 
Astron. J. {\bf129}, 1428 (2005).
\bibitem[40]{40} V. Belokurov, D. Erkal, N. W. Evans, et al., 
Monthly Not. Roy. Astron. Soc. {\bf478}, 611 (2018).









\end{thebibliography}
\end{document}